\documentclass[fleqn,usenatbib]{mnras}

\usepackage{newtxtext,newtxmath}

\usepackage[T1]{fontenc}
\usepackage{ae,aecompl}

\makeatletter
\newcommand{\@todonotes@enable}{1}
\newcommand{\@todonotes@inline}{1}
\makeatother
\input{notes.tex}

\usepackage{graphicx}	
\usepackage{amsmath}	
\usepackage[normalem]{ulem}

\newcommand{\vmax}{\ensuremath{v_\mathrm{max}}}
\newcommand{\rmax}{\ensuremath{r_\mathrm{max}}}
\newcommand{\kpc}{\ensuremath{\,\mathrm{kpc}}}
\newcommand{\kms}{\ensuremath{\,\mathrm{km\,s}^{-1}}}

\DeclareMathOperator{\sech}{sech}

\title[Dark Matter Subhalo Evolution]{High-Resolution Simulations of Dark Matter Subhalo Disruption in a Milky Way-like Tidal Field}

\author[Webb \& Bovy]{Jeremy J. Webb$^1$ \thanks{E-mail: webb@astro.utoronto.ca (JW)} \& Jo Bovy$^1$ \\
$^1$Department of Astronomy and Astrophysics, University of Toronto, 50 St. George Street, Toronto, ON, M5S 3H4, Canada \\
}

\date{Accepted XXX. Received YYY; in original form ZZZ}

\pubyear{2020}

\begin{document}
\label{firstpage}
\pagerange{\pageref{firstpage}--\pageref{lastpage}}
\maketitle

\begin{abstract}

We compare the results of high-resolution simulations of individual dark matter subhalos evolving in external tidal fields with and without baryonic bulge and disk components, where the average dark matter particle mass is three orders of magnitude smaller than cosmological zoom-in simulations of galaxy formation. The Via Lactea II simulation is used to setup our initial conditions and provides a basis for our simulations of subhalos in a dark matter-only tidal field, while an observationally motivated model for the Milky Way is used for the tidal field that is comprised of a dark matter halo, a stellar disk, and a stellar bulge. Our simulations indicate that including stellar components in the tidal field results in the number of subhalos in Milky Way-like galaxies being only $65\%$ of what is predicted by $\Lambda$ Cold Dark Matter ($\Lambda$CDM). For subhalos with small pericentres $(r_p \lesssim 25$ kpc), the subhalo abundance is reduced further to $40\%$, with the surviving subhalos being less dense and having a tangentially-anisotropic orbital distribution. Conversely, subhalos with larger pericentres are minimally affected by the inclusion of a stellar component in the tidal field, with the total number of outer subhalos $\approx 75\%$ of the $\Lambda$CDM prediction. The densities of outer subhalos are comparable to predictions from $\Lambda$CDM, with the subhalos having an isotropic distribution of orbits. These ratios are higher than those found in previous studies that include the effects baryonic matter, which are affected by spurious disruption caused by low resolution.

\end{abstract}


\section{Introduction} \label{s_intro}

In the $\Lambda$ Cold Dark Matter ($\Lambda$CDM) framework, dark matter halos are assembled through the mergers and evolution of dark matter subhalos \citep{white91, springel05}. The end result of this hierarchical galaxy formation process is a halo consisting of an underlying smooth distribution of dark matter, which forms out of the dissolution of subhalos, and substructure in the form of surviving dark matter subhalos. Subhalo dissolution can occur through a variety of mechanisms, including internal relaxation, dynamical friction, tidal heating and tidal shocks \citep{ghigna98, moore99, klypin99, Hayashi03, diemand08, Penarrubia08, stadel09, Penarrubia10}. A subhalo is only expected to survive for a Hubble time if it is sufficiently extended and massive enough so internal mechanisms do not lead to significant mass loss, while at the same time being compact enough so it is shielded from the tidal field of the host galaxy. Therefore, the fraction of dark matter that exists in the form of substructure is likely to be higher in regions of the galaxy where the tidal field is weak \citep{diemand08, springel08}. The expected number of massive subhalos ($M>10^9 M_{\odot}$) predicted by $\Lambda$CDM is in agreement with the satellite galaxy population of the Milky Way, once the suppression of star formation due to the reionization of the Universe is accounted for \citep{Bullock02a,Somerville02a,Koposov09a}. Low mass subhalos ($M<10^9 M_{\odot}$), on the other hand, remain dark as they do not form stars. Hence it is difficult to directly test $\Lambda$CDM predictions of the Milky Way's dark matter substructure content.

Even within the $\Lambda$CDM framework, theoretical predictions of what the distribution of subhalo properties should be tend to vary. Early results from dark matter only simulations \citep{diemand08, springel08} form the basis for most $\Lambda$CDM  predictions. They provide estimates for properties like the radial distribution of subhalos, the subhalo mass function, and the subhalo mass-size relation in Milky Way-like galaxies. However the inclusion of baryonic matter in these simulations, combined with the ever increasing resolution capabilities of new cosmological codes, has caused these results to start being challenged.

Simply adding an axisymmetric baryonic component to the external potential alone is expected to increase subhalo dissolution rates, as inner region subhalos will be subjected to disk shocking \citep{Gnedin97}. \citet{DOnghia10} finds that including the effects of disk shocking on subhalo evolution results in $\Lambda$CDM overestimating the number of high mass ($10^9\,M_{\odot}$) subhalos within 30 kpc of the Galactic centre by a factor of 2. Hence even massive subhalos that contain luminous matter, like the satellite galaxies of the Milky Way, are affected by the Milky Way's baryonic component \citep{Zolotov12, Brooks13, Brooks14}. \citet{DOnghia10} also finds that lower mass subhalos ($10^7\,M_{\odot}$) are overestimated by a factor of 3. Including the various physical mechanisms associated with baryonic matter, like star formation and feedback, suggest that dark matter only simulations overestimate the amount of substructure in Milky Way-like galaxies by even higher factors. Recent work by \citet{Sawala17} and \citet{Richings20} find that $\Lambda$CDM can overestimate the number of subhalos in a galaxy by up to a factor of 4 or 5 respectively, primarily in the inner regions of the galaxy.

The effects of baryonic matter on subhalo evolution are also studied in large suites of hydrodynamic cosmological zoom-in simulations of galaxy formation, like the Feedback in Realistic Environments (FIRE) project, which features both dark and baryonic matter particles. Baryonic matter exists in the form of both gas and stars, which allows for the effects of stellar feedback on dark matter subhalos to be explored. \citet{garrisonkimmel17} find that the dark matter substructure fraction in Milky Way-like galaxies is likely lower than predicted by $\Lambda$CDM as stellar feedback accelerates the dissolution of subhalos. Using the `Latte' simulation \citep{wetzel16}, which is part of FIRE, the authors predict that Milky Way-like galaxies should have $50\%$ the number of subhalos within 100 kpc than predicted by $\Lambda$CDM due to baryonic physics. This factor drops to $20\%$ within 25 kpc and zero within 15 kpc for subhalos with masses greater than  $3 \times 10^6 M_\odot$. It is important to note that in the Latte simulation,  dark and baryonic mass particles have masses of $35,000\,M_\odot$ and $7,070\,M_\odot$ respectively and softening lengths of 20 pc and 4 pc, respectively. Furthermore, only collections of dark matter particles with $\vmax$ greater than $5\kms$ are considered to be subhalos.

\citet{garrisonkimmel17} further find that similar results can be obtained by simply embedding an analytic disk potential that grows with time in a dark matter-only simulation. Motivated by this result, the ELVIS project \citep{kelley19} re-ran cosmological dark matter only zoom-in simulations of Milky Way-like galaxies with an embedded stellar disk, gaseous disk, and bulge component. In the ELVIS suite of simulations, dark matter particles have masses of $30,000\,M_\odot$ and softening lengths of 37 pc. \citet{kelley19} find that subhalos with small pericentres are completely destroyed, suggesting no subhalos with maximum circular velocities ($\vmax$) greater than $4.5\kms$ (the lower limit considered by \citep{kelley19}) should exist within 20 kpc of the Galactic centre and that the substructure fraction within 20 kpc has been low for the past 8 Gyr. Thus, both the FIRE and ELVIS suites of simulations suggest that dark-matter-only $\Lambda$CDM simulations overestimate the amount of dark matter that exists as substructure in the Milky Way by a significant amount.

The limiting factor of both the FIRE and ELVIS suites of simulations is the resolution and softening length of their dark matter particles, which are $3 \times 10^4\,M_\odot$  and $3.5\times 10^4\,M_\odot$  respectively. More specifically, a $3\times 10^6\,M_\odot$ subhalo consisting of 100 $3\times 10^4\,M_\odot$ dark matter particles will evolve differently than a $3 \times 10^6\,M_\odot$ subhalo consisting of three million $1\,M_\odot$ dark matter particles, because an increase in the particle number increases the subhalo's dissolution time due to relaxation. Decreasing the softening length of dark matter particles to match the decrease in their mass will also increase the subhalo's self gravity, allowing it to remain bound for a longer time when subject to an external tidal field \citep{vandenbosch18a}. Simulations by \citet{vandenbosch18b} found that when the mass resolution of dark matter particles was increased and their softening lengths were decreased, most subhalos survived in some capacity to reach a redshift of zero. The results of \citet{vandenbosch18b} therefore suggest that most cosmological simulations that are limited by the force and mass resolution of their dark matter particles will overestimate subhalo disruption rates. Complicating matters further is that $\Lambda$CDM is but one model for dark matter, with warm dark matter, self-interacting dark matter, and 'fuzzy' dark matter (among others) each yielding their own prediction of how the dark matter substructure fraction behaves as a function of galactocentric distance \citep[e.g.][]{press90, hu00, spergel00, vogelsberger12, elbert15, ludlow16, hui17}.

Observational searches for dark matter substructure, which are centred around detecting dark matter through its interactions with light and baryonic matter, have been unable to strongly rule out any of the theoretical estimates above. Gravitational lensing offers one method for constraining the properties of dark matter subhalos, because substructure leads to anomalies in lensed objects that would not occur if the gravitational lens itself was a smooth distribution of dark matter \citep{mao98,gilman20}. Within the Milky Way, the properties of stellar streams are commonly used to constrain the dark matter substructure fraction in the Galaxy and the subhalo mass function. With prominent stellar streams spanning tens of degrees across the sky, they have a high encounter rate with subhalos relative to other systems like star clusters. Interactions with subhalos are predicted to lead to underdensities \citep{bovy17} and perhaps spurs of stars \citep{bonaca19} along a stellar stream. Hence many studies have used the detailed properties of stellar streams to estimate the substructure fraction of the Milky Way and the subhalo mass function \citep{yoon11, carlberg12, erkal15, erkal15b, bovy16, carlberg16,banik19, bonaca19}. Using the Pal 5 stream, \citet{banik19} accounted for how baryonic and dark matter substructure would perturb the system and determined that dark matter substructure was not needed to produce the observed underdensities and overdensities. In a more recent study, \citet{banik19b} performed a similar analysis on both Pal 5 and the GD-1 stream and concluded that a population of subhalos with masses between $10^7 M_\odot$ and $10^9 M_\odot$ are necessary to reproduce the observed properties of both streams. More specifically, their results indicate that the Milky Way contains approximately $20$ to $40\%$ as many subhalos than predicted by dark-matter-only $\Lambda$CDM simulations, which corresponds to a substructure fraction of 0.14 $\%$. Given that both Pal 5 and GD-1 are within 20 kpc of the Galactic centre, this substructure fraction estimate is higher than the fraction predicted by the FIRE and ELVIS suite of cosmological simulations.

The discrepancy between the observational constraints on dark matter substructure in \citet{banik19b} and theoretical predictions made by \citet{garrisonkimmel17} and \citet{kelley19} could potentially be alleviated by increasing the mass solution and decreasing the softening length of dark matter particles in the FIRE and ELVIS suites of simulations. As previously discussed, such a change is expected to increase the dissolution time of individual subhalos \citep{vandenbosch18a, vandenbosch18b}. \citet{vandenbosch18b} finds that in order to accurately model the evolution of a dark matter subhalo in an external tidal field, such that its evolution is not affected by discreteness noise due to poor mass resolution or inadequate force softening, the following conditions must be satisfied:
\begin{equation}\label{eqn:nres}
f_\mathrm{bound} > 0.32 (N_\mathrm{acc}/1000)^{-0.8}
\end{equation}
and 
\begin{equation}\label{eqn:eres}
f_\mathrm{bound} > 1.12 \frac{c^{1.26}}{f^2(c)} \left(\frac{\epsilon}{r_{s,0}}\right)^2
\end{equation}
where $f_\mathrm{bound}$ is the fraction of initial dark matter particles still bound to the subhalo, $N_\mathrm{acc}$ is the number of particles in the subhalo when it is first accreted by a central galaxy, c is the NFW concentration parameter \citep{Navarro96}, $r_{s,0}$ is the initial NFW scale radius, and $\epsilon$ is the dark matter particle softening length. $f(c)$ is given in \citet{vandenbosch18b} to be $\ln(1+c)-c/(1+c)$. However, performing cosmological zoom-in simulations of Milky Way-like galaxies that meet these requirements would require a factor of 1000 increase in the number of dark matter particles if a $10^6 M_\odot$ subhalo is to be followed to $1\%$ of its initial mass. 

An alternative approach, which we pursue here, is to simulate the evolution of individual subhalos in an external field with an embedded stellar component. As previously discussed, \citet{garrisonkimmel17} found that treating the baryonic component of a galaxy as an embedded analytic potential yielded similar dark subhalo populations as more complex simulations that include baryonic particles, indicating that the main baryonic effect on subhalo disruption is the stronger tidal field rather than details of stellar feedback and other baryonic processes. Modelling individual subhalos allows for very high mass resolution and small particle softening, with the only computational expense being the number of subhalo simulations required to represent the subhalo population of the Milky Way. This approach was adopted by \citet{Errani17}, who performed high resolution simulations of massive ($M > 10^8\,M_\odot$) subhalos in an external tidal field containing a disk. The authors find that the presence of a disk results in a factor of two fewer subhalos with $r_p <$ 20 kpc compared to a dark-matter only tidal field, with cored subhalos being more easily disrupted than cuspy subhalos . In this study, we focus on the evolution of subhalos with masses down to $10^6\,M_\odot$ in Milky Way-like tidal fields.


At a redshift 0, the publicly available Via Lactea II (VL2) simulation \citep{diemand07, diemand08} contains 8,247 subhalos within the virial radius $r_v=295.6\,\mathrm{kpc}$ of the Milky Way's NFW halo (here we assume the Milky Way's potential to be the \texttt{MWPotential2014} model from \citep{bovy15}). Only 6,591 of those have pericentres within 100 kpc of the Galactic centre, which will be the subhalos that are primarily responsible for disrupting stellar streams. Since subhalos can be treated as collisionless systems, it is not computationally expensive to simulate the evolution of individual subhalos from VL2 in external tidal fields with and without an embedded stellar component using the criteria from \citet{vandenbosch18b}. From these simulations, we will be able to conclude whether or not $\Lambda$CDM overestimates the substructure fraction of Milky Way-like galaxies and by how much without dark matter particle resolution affecting the result.

In Section \ref{s_method} we describe our method for initializing and simulating a large suite of dark matter subhalos in analytic external fields with and without an embedded stellar component. In Section \ref{s_results} the results of the simulations are presented, with particular focus on how the mass and size of individual subhalos evolve from their time of infall to redshift zero. We will directly compare subhalos simulated in a dark matter only potential and a potential containing a stellar bulge, stellar disk, and the same dark matter halo. We further explore in Section \ref{s_discussion} by how much $\Lambda$CDM overestimates the substructure of the Milky Way, trends that exist between galactocentric distance and the substructure fraction, subhalo mass, subhalo size, and the distribution subhalo orbits. We summarize our findings in Section \ref{s_conclusion}.

\section{Method} \label{s_method}

\subsection{Initial conditions from Via Lactea}

The purpose of this study is to perform high resolution simulations of how dark matter subhalos evolve in potentials with and without a stellar component to determine how including a stellar component affects subhalo dissolution. In setting up the initial conditions of our large suite of simulations, we make use of publicly available data from the VL2 simulation \citep{diemand07, diemand08}. The purpose of the Via Lactea Project was to perform simulations of how Milky Way-like dark matter halos form, which then provide estimates of the present day distribution of subhalo positions, velocities, and masses. At a redshift of zero, there exists 20,048 halos and subhalos in VL2 that have reached a peak circular velocity larger than 4\kms\ at some point in the simulation. Smaller systems do exist but are poorly resolved. The dark matter-only potential of the central host Milky Way-like galaxy at redshift zero was also studied by \citet{diemand08} and can be fit with either a cored \citep{navarro04} and a cusped dark matter density profile \citep{diemand04}. While the evolutionary track of each subhalo is provided, such that its properties at the time it was accreted by the central galaxy are known, a direct re-simulation of each individual subhalo in VL2 is not possible as the time evolution of the central halo's potential is not analytic (which we require in order to perform high resolutions of individual subhalos). Furthermore, any re-simulations of the VL2 could not be directly compared to observational constraints placed on subhalos in the Milky Way as our Galaxy's dark matter halo is not exactly the same as the central host in VL2. Therefore, as described in the following subsections, slight modifications are made to the potential and subhalos in VL2 in order to resimulate each subhalo in an observationally-motivated Milky Way-like potential.

\subsection{Gravitational potentials}

The \texttt{MWPotential2014} galaxy model from \citet{bovy15} is an observationally-motivated potential that accurately reflects the distribution of matter in the Milky Way. The stellar components of the potential are a spherical power-law bulge and a a \citet{Miyamoto75} disc. The dark matter component of the potential, which is an NFW halo \citep{Navarro96} with a virial radius and mass of 295.6\kpc\ and $8.8 \times 10^{11}\,M_\odot$ (defined here as where the mean density is 200 times the mean matter density, assuming $H_0 = 70\kms\,\mathrm{Mpc}^{-1}$ and $\Omega_m = 0.3$), is qualitatively similar to the dark matter potential of the central host galaxy in VL2. The best-fit cusped dark matter density profile \citep{diemand04,diemand08} to the central VL2 galaxy only needs to be scaled by a factor of 1.27 in radius and 2 in mass in order to be equal to the dark matter component of \texttt{MWPotential2014}. By scaling the positions and velocities of subhalos in VL2 by 1.27, we can therefore initiate the subhalo population that $\Lambda$CDM would predict to exist in \texttt{MWPotential2014} and the dark-matter-only potential within which we re-simulate subhalos is then the NFW potential from \texttt{MWPotential2014} (hereafter referred to as the NFW model). High resolution simulations of subhalos in \texttt{MWPotential2014} (hereafter referred to as the MW model) will then demonstrate the effect that baryonic matter has on the evolution of dark matter subhalos.

It is important to note that in both VL2 and ELVIS (discussed above), both the dark matter halo and the stellar component of the central galaxy grow over time. In the setup described above, both components remain fixed in time. This assumption is necessary as the potential of the central VL2 halo does not evolve analytically with time. Only its maximum circular velocity, radius of maximal circular velocity, tidal radius $r_t$, and mass within $r_t$ are given at earlier times. Thus, our simulations of subhalos in the static NFW model will slightly overestimate their dissolution rate with respect to the $\Lambda$CDM expectation. However, the dissolution of subhalos in the NFW model and the MW model will be equally affected by the presence of the halo's tidal field that is constant in time, such that their relative differences will reflect how the distribution of subhalos in the Milky Way and their properties will differ from $\Lambda$CDM predictions in the presence of a massive disk. Similarly, the disk's mass and radial profile is fixed and the disk's tidal field at earlier times is therefore somewhat larger than it should be in reality; the effect of the disk on subhalo disruption will therefore be overestimated, although by only a small amount as much of the disruption happens at epochs when the disk's mass is mostly in place.

\begin{figure}
    \includegraphics[width=0.48\textwidth]{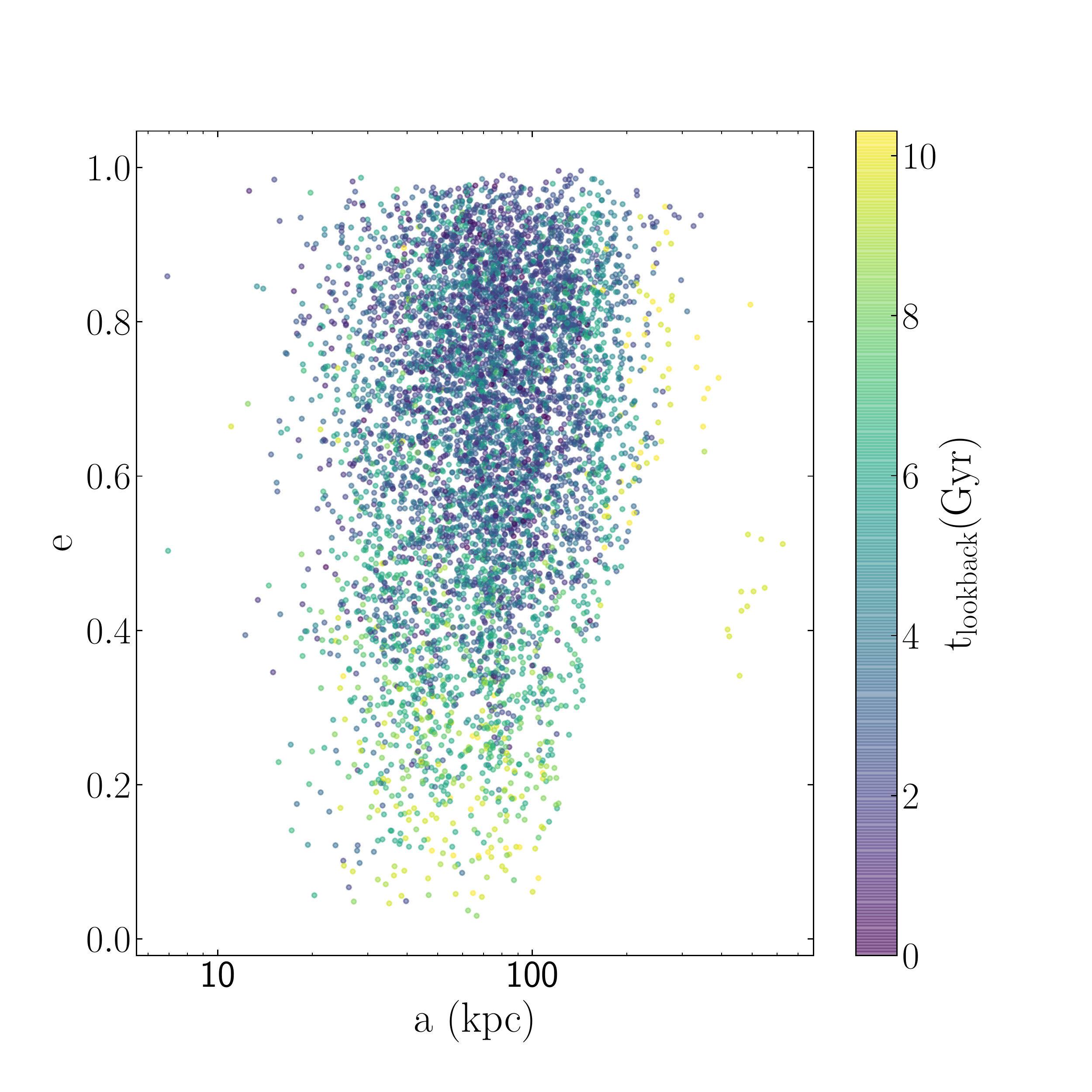}
    \caption{Semi-major axis and orbital eccentricity of VL2 subhalos to be re-simulated at high resolution, colour coded by their infall time.}
    \label{fig:via_lactea}
\end{figure}

\subsection{Subhalos}

Given the positions and velocities of subhalos in VL2 at redshift zero, we first integrate their orbits using \texttt{galpy} \footnote{http://github.com/jobovy/galpy} \citep{bovy15} in the cusped dark matter density profile that \citet{diemand08} fits to the central Milky Way-like halo. The purpose of this integration is to determine each subhalo's infall time, where we define infall time as the moment when a subhalo's distance from the Galactic centre becomes equal to or less than its present day apocentre. The distance of each subhalo as a function of redshift is provided by VL2, however it should be noted that we scale these distances based on how much the tidal radius of the central host galaxy changes with redshift to account for the central galaxy's growth. Due to this additional scaling, we only consider subhalos that have infall times less than or equal to a redshift of 3. Prior to a redshift of 3 the central halo in VL2 is not consistently found by VL2's halo finding scheme \citep{diemand06,diemand07}. Determining each subhalo's infall time provides us with an estimate for how long each subhalo as been evolving under the influence of the central galaxy's tidal field. 

While VL2 also provides the position and velocity of each subhalo at infall, their orbital properties reflect the potential of the central Milky Way-like halo at their infall and not its current potential. Hence using these positions and velocities with the static potential at a redshift of zero to initialize subhalo orbits would result in subhalos being subjected to a different external tidal field than the one they experience during their long-term evolution. Therefore, while we use each subhalo's mass at infall as its initial mass, its orbital position and velocity at redshift zero are used to initialize subhalo orbits. This approach is valid given that we use a static potential emulating the redshift zero mass distribution to evolve the orbtis and given that the purpose of using data from VL2 is simply to initialize a subhalo population with a distribution of masses and orbits that accurately reflects the predictions of $\Lambda$CDM, and not re-simulate VL2 subhalos specifically.

It should be noted that the resolution of VL2 will result in subhalos artificially disrupting on shorter timescales \citep{vandenbosch18a}. Hence the subhalo population at a redshift of zero is incomplete. However, because we are mainly interested here in determining the relative ratio of the number of surviving subhalos in a Milky Way-like halo to that in the dark-matter-only halo, this incompleteness only affects our results if specific combinations of subhalo mass and orbit are preferentially disrupted due to resolution effects, which given Equations \ref{eqn:nres} and \ref{eqn:eres} appears unlikely since the primary factors are the mass and softening lengths of dark matter particles and the density profile of the subhalo.

In order to properly re-simulate subhalos from VL2 in the MW and NFW potential models, it is necessary to scale their positions and velocities in VL2 at redshift zero by a factor of 1.27. As discussed above, this scaling is necessary to account for the slight difference between the cusped dark matter density profile fit to the central Milky Way-like halo potential in VL2 and the NFW-halo dark matter component of \texttt{MWPotential2014}. The scaling ensures that the distribution of subhalo orbital properties, like pericentre and semi-major axis, are conserved.

\begin{figure}
    \includegraphics[width=0.48\textwidth]{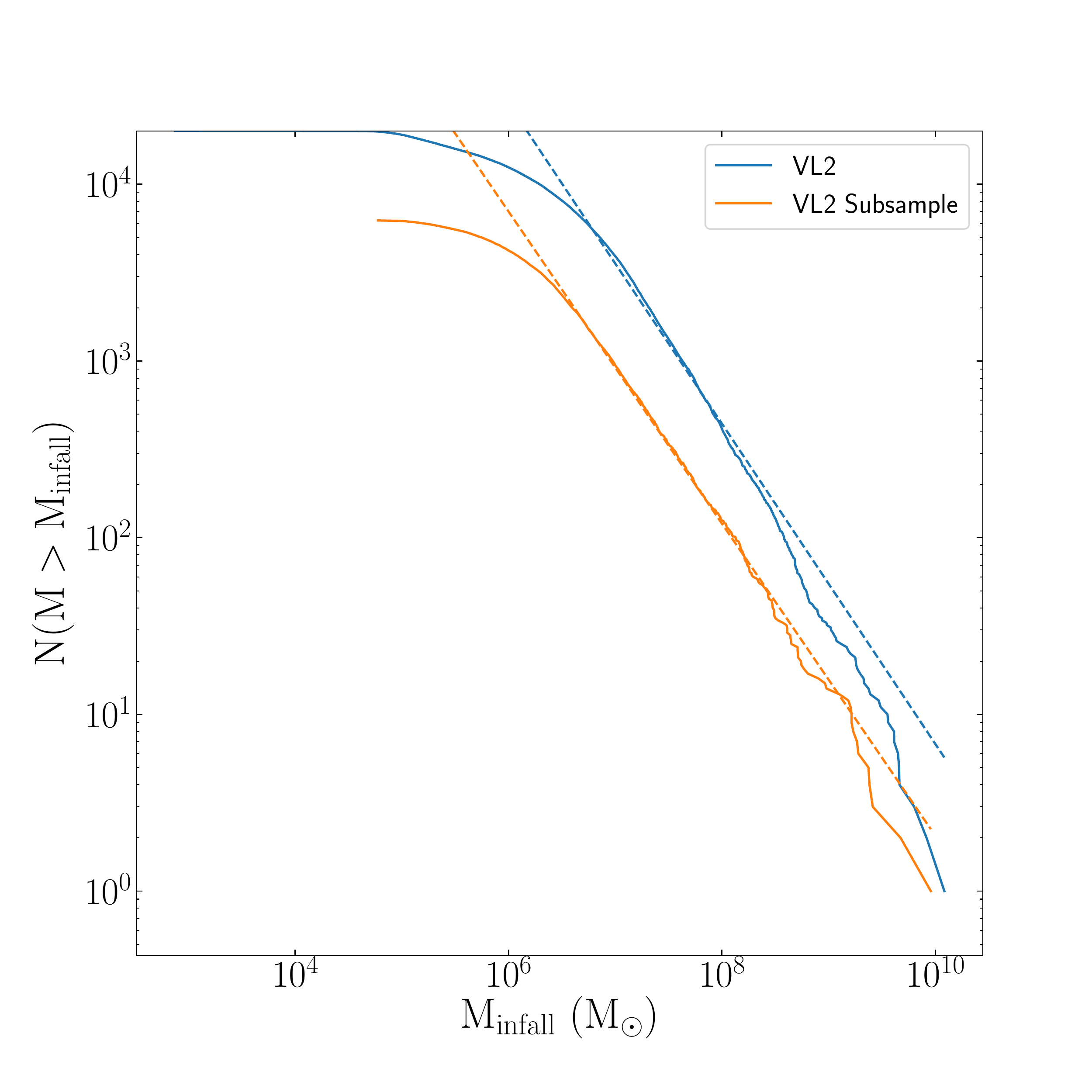}
    \caption{Cumulative mass functions of subhalo masses at infall for the entire VL2 population (blue) and the subset of subhalos that we re-simulate at high resolution (orange). Beyond $10.0^{6.5}$, both mass functions are well fit by power laws of the form $N \propto M^{-0.9}$.}
    \label{fig:via_lactea_imf}
\end{figure}

\begin{figure}
    \includegraphics[width=0.48\textwidth]{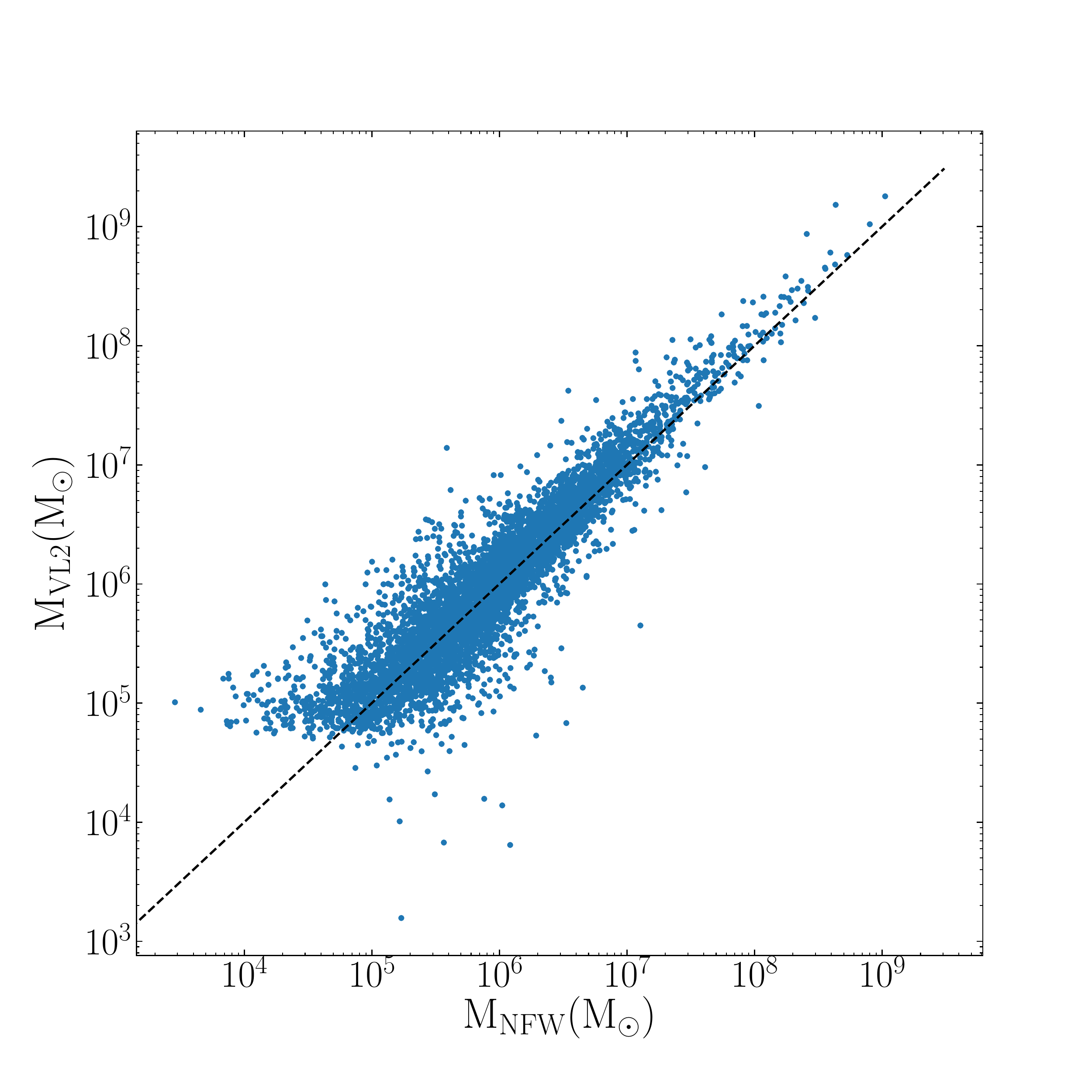}
    \caption{Mass of subhalos from the VL2 simulation compared to the same initial subhalo evolved in the NFW potential that survive to reach a redshfit of zero. The black dashed line illustrates a 1:1 relationship.}
    \label{fig:via_lactea_nfw_compare}
\end{figure}

\begin{figure}
    \includegraphics[width=0.48\textwidth]{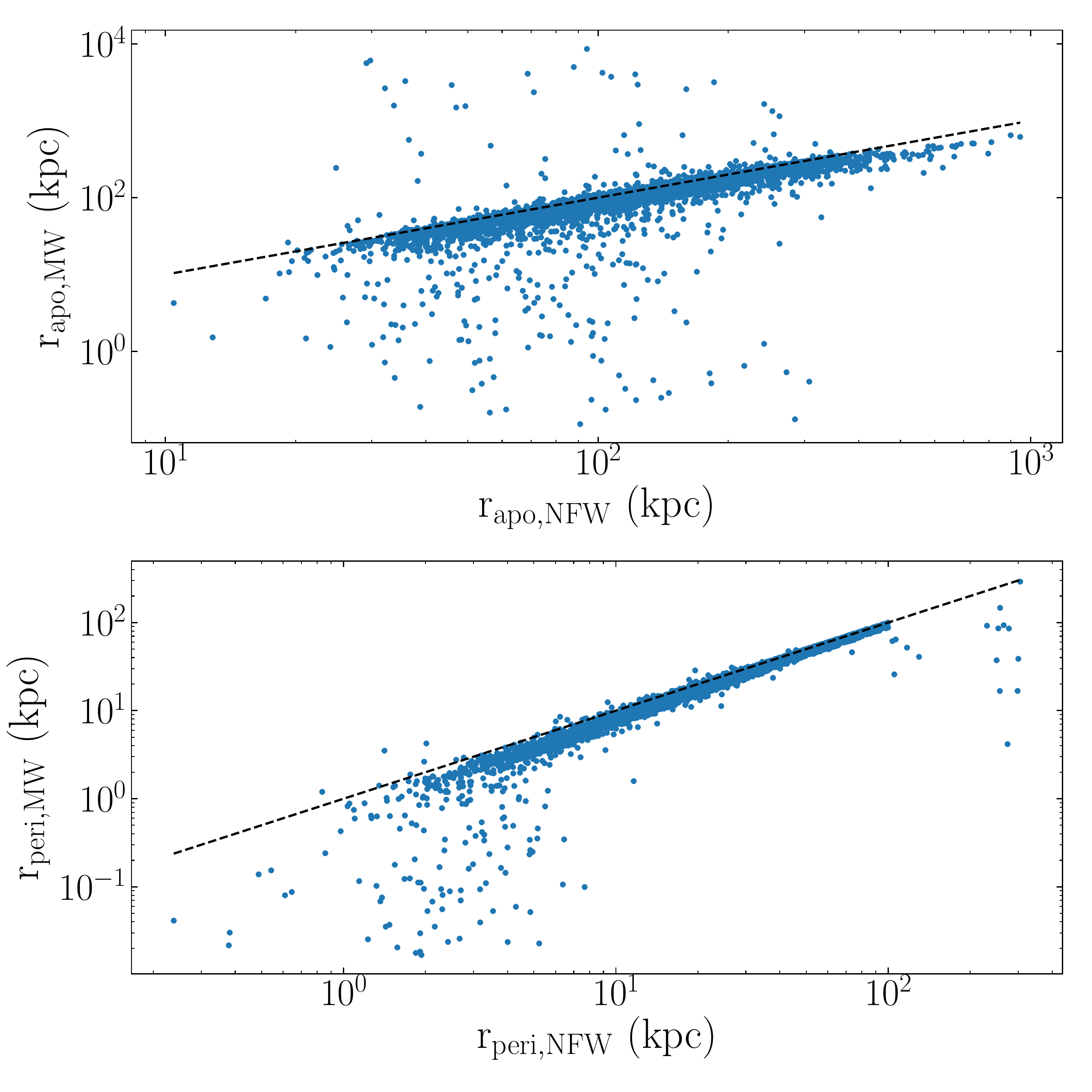}
    \caption{Comparison of subhalo apocentres (top panel) and pericentre (bottom panel) in the NFW and MW models}.
    \label{fig:rp_rap_compare}
\end{figure}

\begin{figure}
    \includegraphics[width=0.48\textwidth]{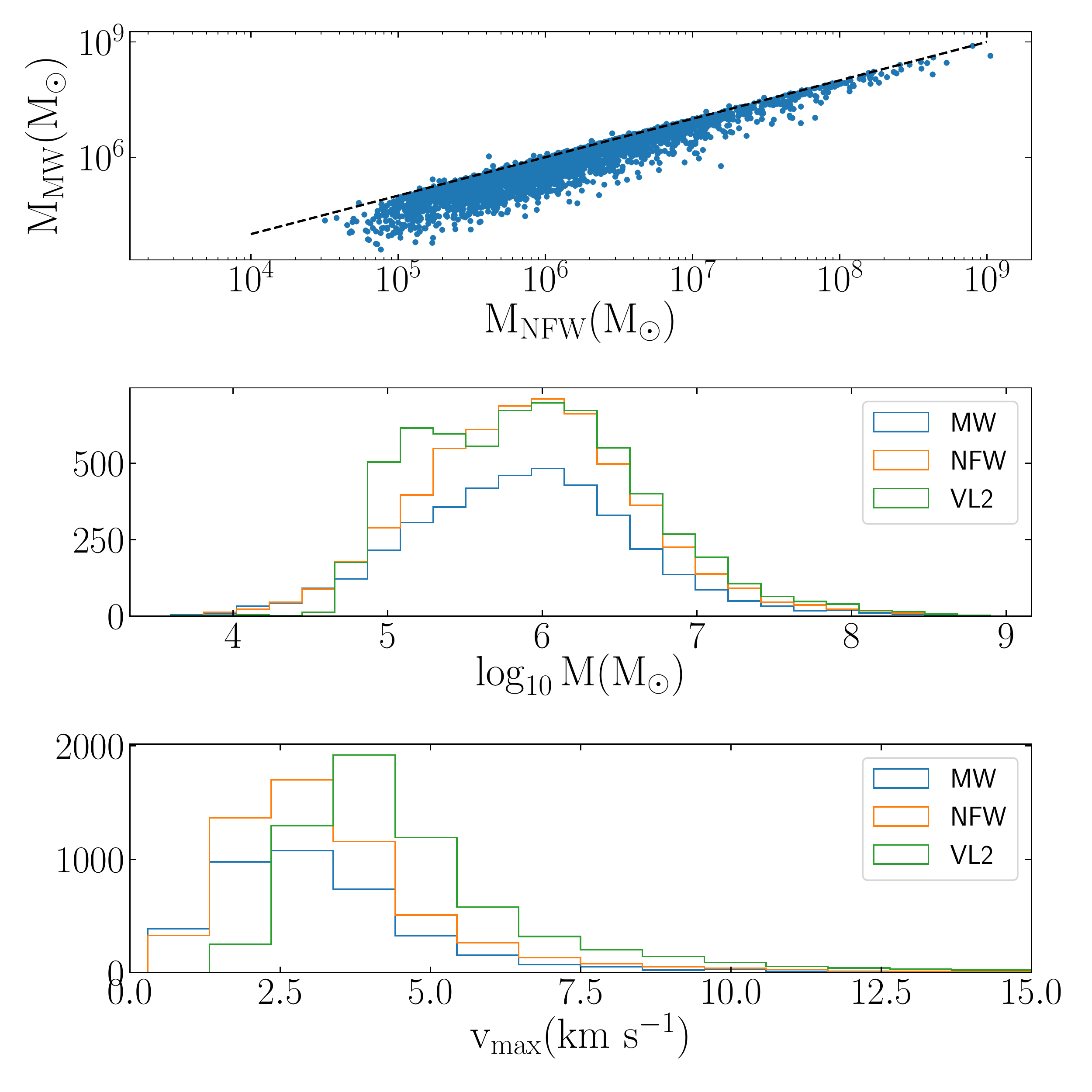}
    \caption{Top Panel: Mass of subhalos evolved in the NFW potential compared to the same initial subhalo evolved in the MW potential that survive to reach a redshfit of zero. The one-to-one relation is given by the dashed line. Middle Panel: Distribution of subhalo masses at redshift zero in the MW potential (blue), the NFW potential (orange) and in VL2 (green). Bottom Panel: Distribution of subhalo maximum circular velocities at redshift zero in the MW potential (blue), the NFW potential (orange) and in VL2 (green). Essentially all subhalos lose more mass in the MW potential than in the dark-matter-only NFW potential, with the distribution of subhalo masses in the NFW potential matching the VL2 subhalo population for $\mathrm M > 10^{6} M_{\odot}$. However subhalos evolved in the NFW potential have lower densities at a redshift of zero than subhalos of comparable mass from VL2.}
    \label{fig:via_lactea_deltam}
\end{figure}

Because we are primarily concerned with comparing the distribution of subhalo properties in our simulations to observational constraints placed on subhalos in the Milky Way using stellar streams, it is not necessary to perform a high resolution simulation of every subhalo in VL2. We specifically only want to consider subhalos that:

\begin{itemize}
    \item are associated with the central Milky Way-like halo in VL2 
    \item do not overlap with another subhalo
    \item have pericentres within 100 kpc of the model galaxy's virial radius 
    \item reach their current orbit after a redshift of 3 
\end{itemize}

\noindent Subhalos with negative values of \rmax\ at infall, which is an indication that they overlapped with a different, larger halo, can be ignored. Additionally, subhalos with pericentres beyond 100 kpc are not considered as they will have a negligible effect on the evolution of observed stellar streams. As discussed above the central Milky Way-like halo in VL2 does not form until a redshift of 3, hence modelling the evolution of subhalos with infall redshifts greater than 3 is unnecessary. 

Combing all four of the above factors, we only end up resimulating the evolution of 6,591 subhalos from their infall time to present day in both the NFW and MW galaxy models. The exact NFW density profile of each subhalo at infall is set by its scale radius $r_s$, tidal radius $r_t$, and the mass within the tidal radius $M_t$ as given by VL2. Note that the central density $\rho_s$ and $r_s$ are calculated as $\rho_s=\vmax/[1.64 r_s]^2/G$ and $r_s =\rmax/2.16258$ \citep{Bullock01} as VL2 only provides $\vmax$ and $\rmax$.

Each subhalo is initialized using the \texttt{mkhalo} routine in \texttt{NEMO} \citep{teuben95, mcmillan07} with a truncation factor equal to $2/(\sech(r/r_t)+1/\sech(r/r_t))$. Note that since \texttt{mkhalo} generates stars out to $10 r_t$, the mass within $10 r_t$ is used for the mass parameter to ensure the mass within $r_t$ is equal to $M_t$. The number of particles and softening length of the particles are set to satisfy Equations \eqref{eqn:nres} and \eqref{eqn:eres}, assuming we want to resolve each subhalos evolution down to $1\%$ of its initial mass. Setting $f_\mathrm{bound} = 0.01$ in Equation \eqref{eqn:nres} results in the subhalo consisting of 988211 particles, with individual dark matter particles having masses of $M_t$/988211. Since we generate particles out to $10 r_t$, there are always more than 988221 particles in a given simulation. Given the scale radius of the subhalo and its NFW central concentration, it is then possible to solve for the softening length $\epsilon$ in Equation \eqref{eqn:eres}. Softening lengths are typically on the order of 0.1 pc and dark matter particles have masses on the order of $10\,M_\odot$. The initial position and velocity of each subhalo are set equal to their scaled redshift zero coordinates in VL2 as we are only interested in how subhalos evolve as a function of orbit. 

\begin{figure*}
    \includegraphics[width=\textwidth]{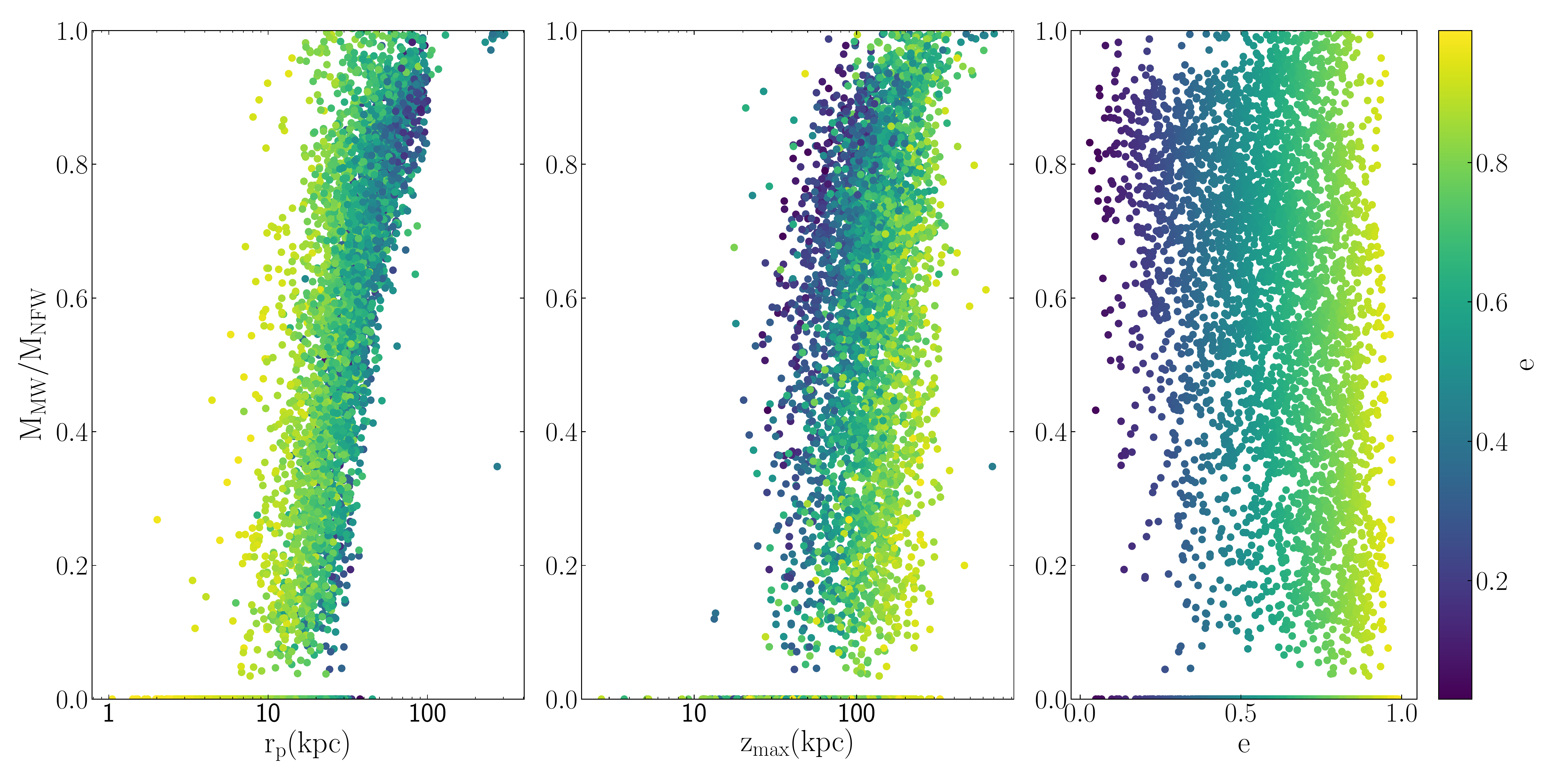}

    \caption{Ratio of the mass in the MW potential to that in the NFW potential for subhalos that survive to reach a redshfit of zero compared to their pericentres (left), maximum heights above the disc (centre), and orbital eccentricity (right) in the NFW potential. Subhalos on orbits with smaller pericentres, with smaller $z_\mathrm{max}$, and larger eccentricities typically lose more mass in the MW potential than those with larger pericentres and $z_\mathrm{max}$ and smaller eccentricities.}
    \label{fig:mrat_orbit}
\end{figure*}

\begin{figure}
    \includegraphics[width=0.48\textwidth]{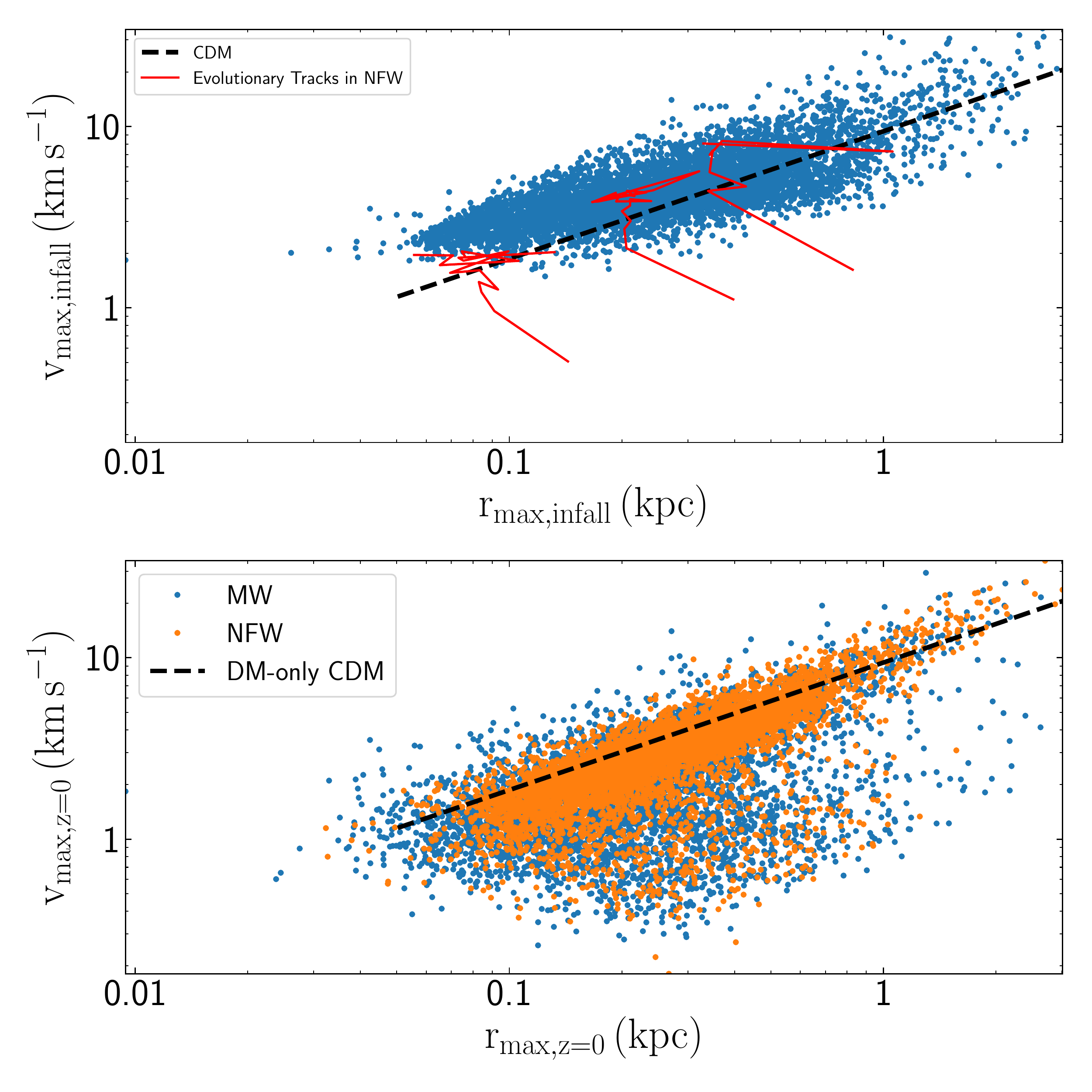}
    \caption{Radius of maximal circular velocity versus maximum circular velocity at infall (top panel) and redshift zero (bottom panel) for resimulated VL2 subhalos. In the top panel, mean evolutionary tracks for subhalos in the NFW model that lose more than $95\%$ of their initial mass are shown in red. In the bottom panel, subhalos in the MW and NFW galaxy models are marked in blue and orange respectively. The standard redshift zero CDM relationship from \citet{Dutton14a} is illustrated as a dashed line. At infall subhalos are above the $z=0$ CDM relation and then evolve towards it, with the subhalos falling far below the $z=0$ CDM relation being those that are approaching full dissolution.}
    \label{fig:via_lactea_rvmax}
\end{figure}

The initial orbit and mass at infall of all simulated subhalos are illustrated in Figures \ref{fig:via_lactea} and \ref{fig:via_lactea_imf}, with the initial subhalo mass function being well fit by a power-law of the form $N(M>M_\mathrm{infall}) \propto M^{-0.9}$ for subhalos between $10^{6.5}$ and $10^{10}\,M_\odot$. Given that the mass function should increase with decreasing mass, Figure \ref{fig:via_lactea_imf} further illustrates that the VL2 subhalo population is incomplete below $10^{6} M_\odot$. Note here, initial mass means every subhalo's mass at its respective infall time. At redshift zero, CDM predicts the slope of the mass function to also be $\approx -0.9$ \citep[e.g.][]{springel08}. Hence the slope of the subhalo mass function remains fairly constant in time.

Once initialized, each subhalo is then evolved for a time equal to the difference between their infall time and the present day in both the NFW and MW galaxy models using the \texttt{GYRFALCON} $N$-body code \citep{dehnen00,dehnen02} within NEMO \citep{teuben95}. At each time step, we estimate the limiting radius of each subhalo by determining where the subhalo's density profile equals the local density of the external tidal field. We then consider the subhalo to be all mass within the limiting radius, with any particles beyond the limiting radius assumed to no longer be part of the subhalo.
 
\section{Evolution of individual subhalos}\label{s_results}

The key metric for comparing our high-resolution simulations of subhalos evolving in an NFW dark-matter-only potential and a realistic MW potential to the evolution of subhalos in VL2 is subhalo mass. The mass loss history of each subhalo in the NFW potential will be slightly different than subhalos in VL2 for several reasons. First, the subhalos are being simulated with much higher mass resolution, which results in subhalos being significantly more difficult to disrupt via external tides \citep{vandenbosch18a,vandenbosch18b}. Second, the NFW potential in our simulations remains constant with time, while in VL2 the central halo is built up via hierarchical formation. This second point alone would lead to a higher subhalo disruption rate as subhalos will be subjected to a stronger mean tidal field over the course of their entire orbit. However, it is important to remember that the NFW galaxy model is slightly less massive than the central VL2 halo, which would actually lead to lower subhalo disruption rates. The combined effects of these three factors are illustrated in Figure \ref{fig:via_lactea_nfw_compare}, where we compare the redshift zero masses of our simulated subhalos to those in VL2. While to first order the two masses are equal, there is significant scatter about the 1:1 relation, with 520 of the 6,591 subhalos completely dissolving in the NFW potential. The general agreement between subhalo masses in the NFW galaxy model and VL2 subhalos at a redshift of zero demonstrates that the subhalo population evolving in the NFW model are an accurate representation of a $\Lambda$CDM subhalo population.

Regardless of the net effect created by these two competing factors, comparing the mass loss history of each subhalo in the NFW potential to those on the same orbit in the MW potential will illustrate how including an axisymmetric stellar component in the external tidal field will affect subhalo evolution. The first effect will be that the orbits of individual subhalos will differ in the NFW and MW due to the additional stellar component. Figure \ref{fig:rp_rap_compare} illustrates the pericentre and apocentre of subhalos in both the NFW and MW models, with the majority of subhalos falling on the 1:1 line. Deviations are only observed at low galactocentric distances where the disk becomes important. With most clusters near the disk having smaller pericentres in the MW model than the NFW model, increased dissolution rates are expected.

In the MW potential, an additional 1,830 subhalos reach complete dissolution. Of the subhalos that survive to reach redshift zero in both models, their masses are compared in the top panel of Figure \ref{fig:via_lactea_deltam}. The middle and bottom panels illustrates the distribution of subhalo masses and maximal circular velocity for each population, with the redshift zero masses of VL2 subhalos in Figure \ref{fig:via_lactea_nfw_compare} also being shown for illustrative purposes. Differences in the dissolution rates of individual subhalos can be attributed to the MW tides being stronger than those in the dark-matter-only NFW potential due to the additional stellar component. The lower $v_\mathrm{max}$ of subhalos in the NFW and MW models, relative to VL2, is also consistent with the work of \citet{Hayashi03}, who finds tidally stripped NFW halos are also heated and become less dense over time.

Taking into consideration the orbital parameters of each subhalo in Figure \ref{fig:mrat_orbit}, subhalos in MW with smaller pericentres $r_p$ are more strongly affected by the presence of the disk and therefore disrupt faster. It should be noted that 41 subhalos have $\rm M_{MW}/M_{NFW}>1$, however these are all subhalos with galactocentric distances larger than 300 kpc. Hence their mass ratio being greater than one can simply be attributed to differences in their orbits caused by the additional stellar component increasing the total mass of the galaxy model. The minimum pericentre of these subhalos is 55 kpc in the MW model and 59 kpc in the NFW model, such that direct interactions between the subhalos and the additional stellar component are not a factor.

\begin{figure}
    \includegraphics[width=0.48\textwidth]{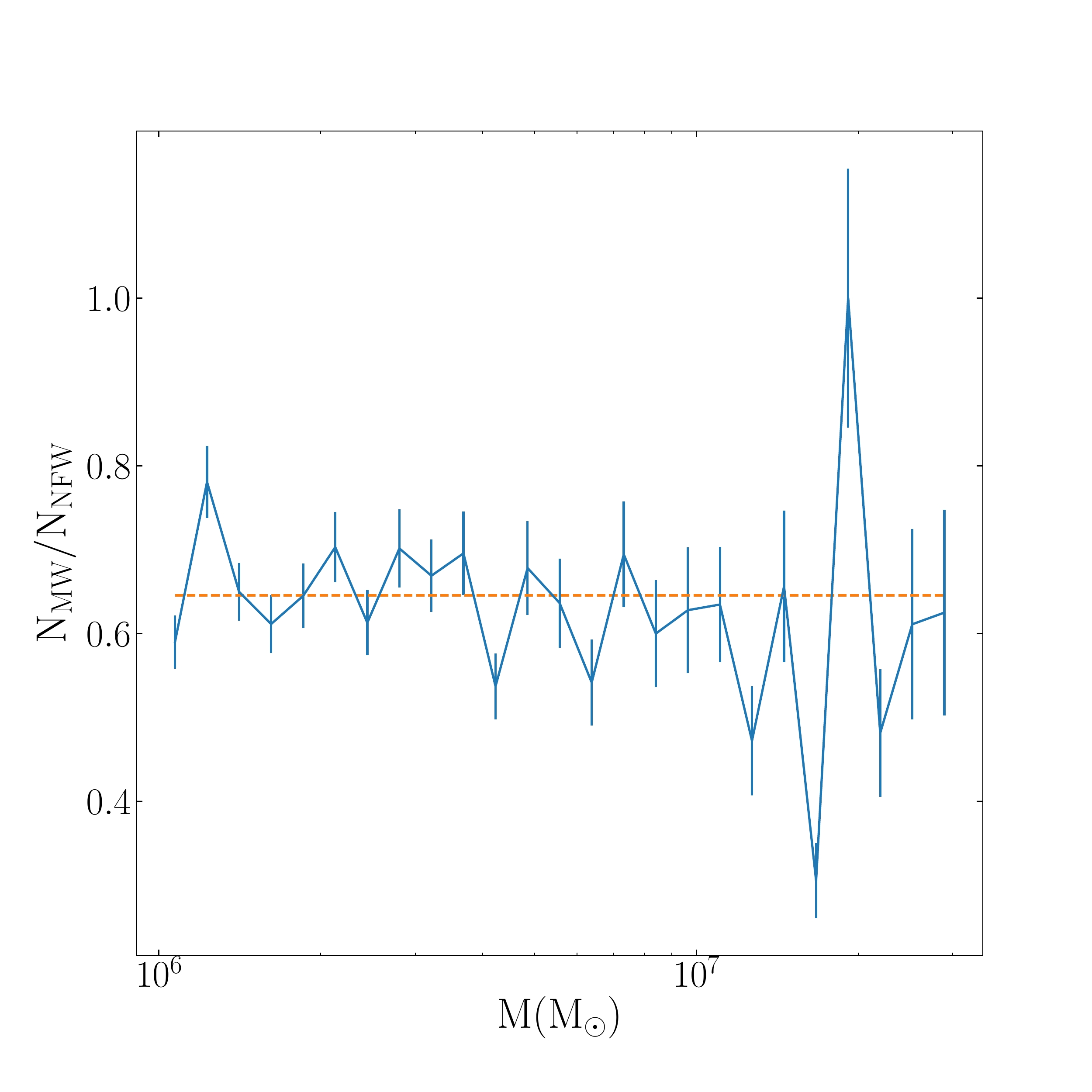}
    \caption{Ratio of number of subhalos in the MW potential to number of subhalos in dark-matter-only NFW potential at redshfit of zero for different mass ranges. The errorbars represent Poisson error. Between $10^6$ and $10^{7.5}\,M_{\odot}$ the effect of the baryonic disk and bulge on the mass function is about 50 to 75\%, with no dependence on mass. The mean ratio for subhalos between $10^{6}$ and $10^{7}\,M_{\odot}$ is $\sim 65\%$ and marked as a horizontal dashed line.}
    \label{fig:nrat}
\end{figure}

\begin{figure}
    \includegraphics[width=0.48\textwidth]{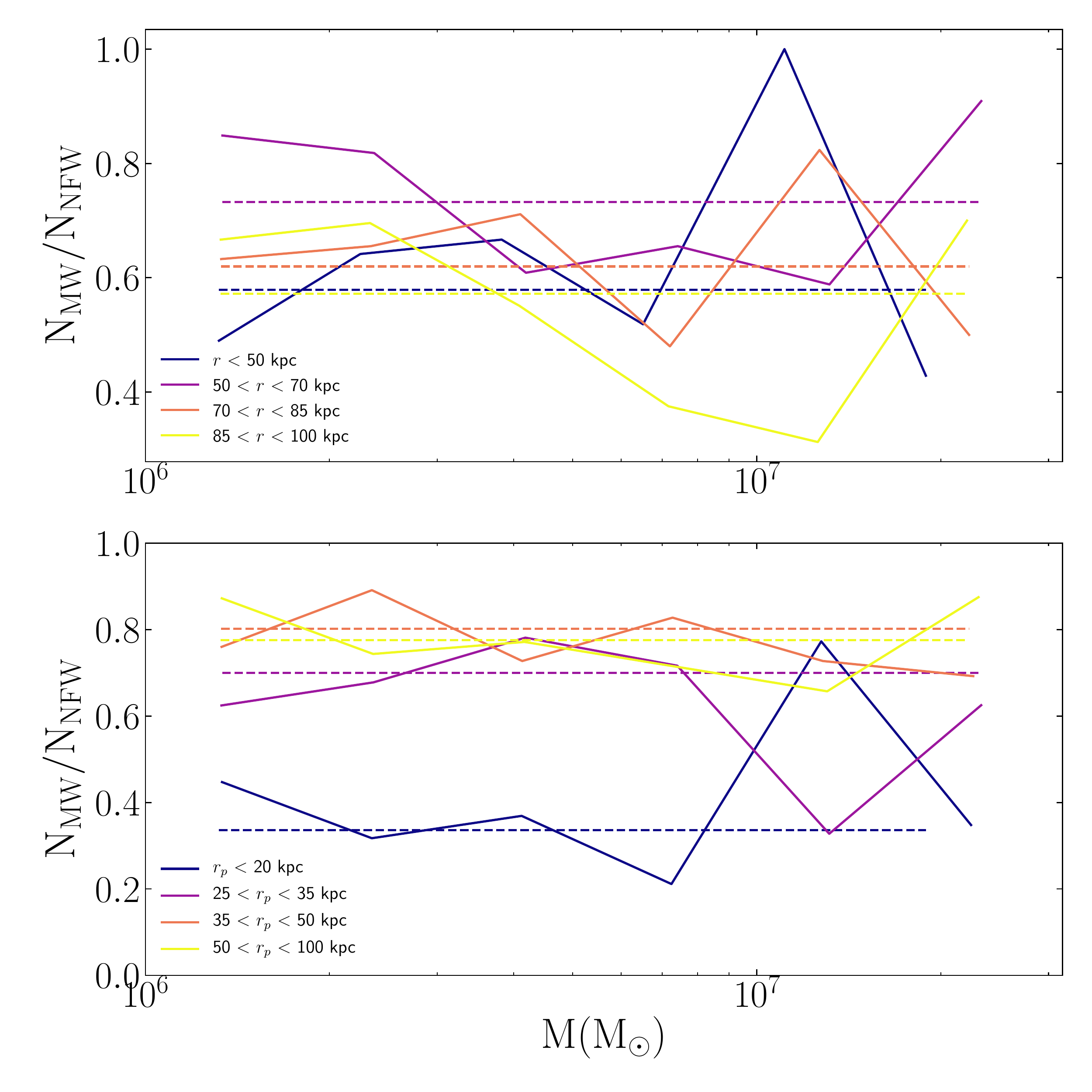}
    \caption{Like Figure~\ref{fig:nrat}, but breaking up the subhalo population into different galactocentric radii (top) and pericentric radii (bottom). Within 100\kpc, the baryonic suppression of the mass function is about 50\%. The abundance of subhalos with pericentres close to the disk and bulge is more strongly suppressed than that of subhalos that remain further away, with a suppression of the subhalo mass function to only 30\% of the subhalo abundance in the dark-matter-only NFW potential for subhalos with $r_p \lesssim 15\kpc$ and $50\%$ for subhalos with $15\kpc \lesssim r_p \lesssim 30\kpc$. The abundance of subhalos that do not venture near the disk or bulge is about 70\% of that in the NFW potential.}
    \label{fig:nrat_rad}
\end{figure}

Given that subhalos reach dissolution faster in the MW potential than in the NFW potential, especially subhalos with orbits that bring them within the Galactic disk and bulge, how the structural evolution of subhalos may depend on the properties of the host galaxy is also of interest. Figure \ref{fig:via_lactea_rvmax} directly compares the radius of maximum circular velocity and maximum circular velocity at redshift zero for subhalos in the MW and NFW potentials, where the standard prediction from CDM from \citet{Dutton14a} is also illustrated. Our simulations indicate that, as expected, subhalos in the NFW potential evolve from their $\rmax$ and $\vmax$ at infall towards the relationship predicted by CDM (see Figure \ref{fig:via_lactea_rvmax}). As illustrated by the mean evolutionary tracks of subhalos that reach at least $95\%$ dissolution, as subhalos lose mass $\vmax$ decreases while $\rmax$ stays relatively constant. Low mass subhalos remain relatively close to the \citet{Dutton14a} relation until they have lost $\sim 75\%$ of their initial mass, while high-mass subhalos remain on the relation until they have lost over $90\%$ of their initial mass. Hence the shape of the density profile of individual subhalos undergoes little evolution, similar to work by \citet{drakos20} on tidally-stripped halos. Similar to the evolution of subhalo masses, there is some scatter about the predicted CDM relationship due to our increased mass resolution and static external potential. The lowest mass subhalos also appear to be a minor exception to this rule, as they appear to expand as they lose mass. This expansion is likely a result of the subhalos approaching complete dissolution, which also explains the small subset of subhalos in the NFW potential that are far below the predicted CDM relationship with values of $\vmax$ that are much lower than CDM would predict given their $\rmax$. 

Together, Figures \ref{fig:via_lactea_deltam} and \ref{fig:via_lactea_rvmax} illustrate that subhalos that evolve in the MW potential have, on average, lower masses and values of $\vmax$ than subhalos that evolve in the NFW potential. Thus, including the bulge and disk results in the subhalos being, on average, less dense than what dark-matter only simulations would suggest. This finding is consistent with subhalos in the MW potential being more easily disrupted. 

\section{Evolution of the subhalo population} \label{s_discussion}

Re-simulating the evolution of VL2 dark matter subhalos with an increased number of particles and decreased softening length in an NFW galaxy model leads to a distribution of subhalo masses at redshift zero that is comparable to VL2. However when a baryonic component is included, namely a bulge and disk, the dissolution rate of individual subhalos increases as we saw in the previous section. The increase can be attributed to both the overall tidal field being stronger and the presence of an axisymmetric component that can shock subhalos that pass through it. We therefore expect that the standard CDM model, refined through cosmological dark matter-only simulations, overestimates the substructure content of Milky Way-like galaxies. In the following subsections, we study the effect of the baryonic component on the population of subhalos, in particular focusing on the effect on the overall mass function, on the internal structure of the subhalos, and on the subhalos' orbital distribution function.

\subsection{Mass}

Figure \ref{fig:nrat} shows the ratio of the subhalo mass functions of all surviving subhalos with masses between $10^6\,M_{\odot}$ and $10^{7.5} M_{\odot}$ in the NFW and MW galaxy models at redshift zero. We purposefully ignore the mass functions below $10^6 M_{\odot}$ as VL2 is incomplete below this limit (see Figure \ref{fig:via_lactea_imf}). Subhalos with masses greater than $10^{7.5} M_{\odot}$ are also ignored as there are very few of them, such that several realizations of their evolution would need to be simulated in order to accurately estimate the high-mass end of the subhalo mass function. Comparing the mass functions of subhalos in both galaxy models, the MW population has fewer subhalos than the NFW potential, as expected from the trend in Figure~\ref{fig:via_lactea_deltam}. For subhalo masses below  $10^{7.5}\,M_{\odot}$, the MW potential typically has between $50\%$ and $75\%$ the number of subhalos as the NFW potential. The mean ratio for subhalos between $10^{6}$ and $10^{7}\,M_{\odot}$ is $\sim 65\%$. Since the ratio is near flat, the cumulative mass distributions of subhalos in both models are nearly identical in shape with a minor offset reflecting the increased number of subhalos in the MW model that reach complete dissolution.

\begin{figure}
    \includegraphics[width=0.48\textwidth]{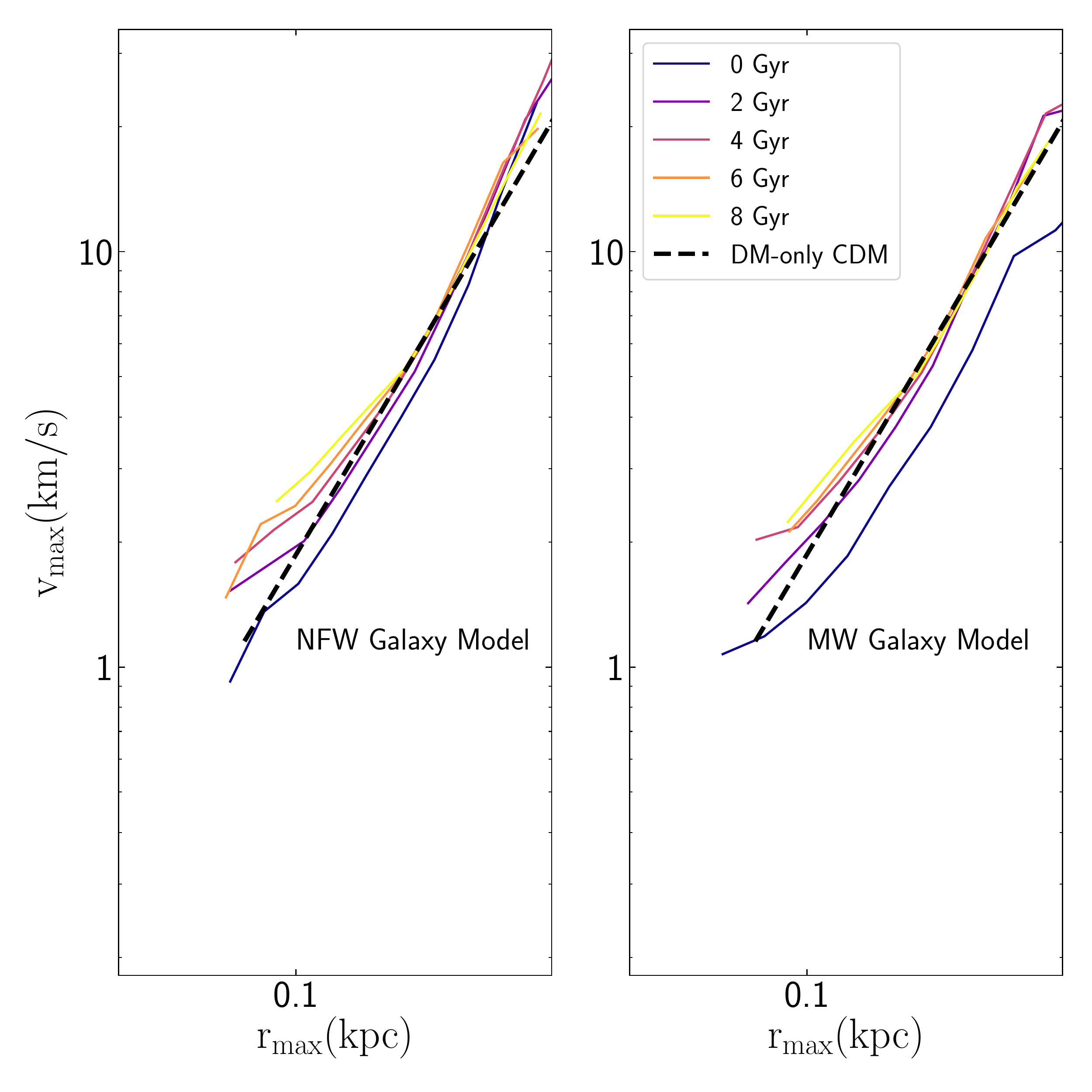}
    \caption{Mean relation between the radius of maximal circular velocity versus maximum circular velocity as a function of lookback time (from 8 Gyr ago to present day for resimulated VL2 subhalos in the MW (left) and NFW (right) galaxy models. The standard redshift zero CDM relationship from \citet{Dutton14a} is illustrated as a dashed line. Uncertainties in the mean $\vmax$ are on the order of $30\%$, with the exception of the $t=0$ Gyr case where the uncertainties are $45\%$. In the Milky Way model, subhalos evolve to have lower density than in the dark-matter-only NFW model.}
    \label{fig:via_lactea_rvmax_evol}
\end{figure}

Looking at the mass function ratios at different galactocentric radii $r$ in Figure \ref{fig:nrat_rad} reveals that the conclusion of the MW potential having between $50\%$ and $75\%$ the number of subhalos as the NFW potential is true within 100 kpc of the galactic centre. However beyond 50 kpc, where the effects of the bulge and disk are much weaker, the factor is closer to $80\%$. To first order it is perhaps surprising that trend with galactocentric distance is not stronger, as the effects of the bulge and disk are significantly stronger within 10 kpc. However it is important to remember that the distribution of subhalo orbits results in many subhalos being located at galactocentric distances that do not reflect the mean or maximum tidal field that they are exposed to. The starkest manifestation of this is the fact that subhalos with $r>85$ kpc have a mean mass function ratio of $\sim 60\%$, comparable to subhalos with $r>50$ kpc. This similarity, and reversal of the trend that mean mass ratio increases with $r$ is due to the orbital anisotropy profile of the subhalo population. The outermost subhalos ($r > 85$ kpc) have a higher degree of radial anisotropy, and therefore a lower median pericentre distance, than subhalos with $70 < r > 85$ kpc. Hence the outer subhalos experience a stronger maximum and mean tidal field and lose more mass. We discuss the orbital anisotropy profile of subhalos in more detail in Section \ref{s_orbits} below.

To better take into account the maximum tidal field that subhalos are subjected to, the bottom panel of Figure~\ref{fig:nrat_rad} breaks up the subhalo population by pericentric radius instead. Now a much clearer trend emerges. Subhalos in the MW potential model with small pericentres are disrupted at a faster rate than in the NFW potential model, resulting in the ratio of number of subhalos in the MW potential to number of subhalos in the NFW potential to be approximately $30\%$ across the full mass spectrum that we consider. For subhalos with larger pericentres that keep them relatively far from the bulge and disk, the ratio is approximately $70\%$ across the full mass spectrum.

\subsection{Structure}

In order to understand the evolution of the subhalo system as a whole, the mean relationship between $\rmax$ and $\vmax$ of subhalos in each galaxy model at different lookback times is shown in Figure \ref{fig:via_lactea_rvmax_evol}. It is important to note that subhalos have a range of infall times (see Figure \ref{fig:via_lactea}), such that at a given time there can be a significant fraction of subhalos that have only recently fallen into the galaxy and have undergone little evolution. Figure \ref{fig:via_lactea_rvmax_evol} reveals that in the NFW potential the mean relationship between $\rmax$ and $\vmax$ does not stray far from the predicted CDM relationship from \citet{Dutton14a}, with a very gradual decrease in the $\vmax$ over time. As previously illustrated in Figure \ref{fig:via_lactea_rvmax}, subhalos do not evolve below the \citet{Dutton14a} until they have undergone significant dissolution. Furthermore, with pristine subhalos falling into the central potential throughout the simulation, it is not until recently that the infall rate has slowed to the point that the subhalo population is dominated by subhalos that have undergone significant dissolution and $\rmax$-$\vmax$ relationship has become offset from the predicted CDM relationship. Conversely, due to the higher disruption rate experienced by sub-halos in the MW potential, subhalos over the entire range of $\vmax$ are less massive than CDM would predict given their $\rmax$.

Next we explore how differences between the structural evolution of subhalos in the MW and NFW potentials may also depend on their location and orbit in the galaxy. We initially find the relationship between $\rmax$ and $\vmax$ of subhalos in each galaxy model at different times and at different locations in the galaxy. More precisely, subhalos are separated based on their current position in the Milky Way and the evolution of the mean relationship between $\rmax$ and $\vmax$ is considered at different times. However, similar to Figure \ref{fig:nrat_rad}, there is no strong relationship between the dissolution of subhalos and their current galactocentric radius.

\begin{figure*}
    \includegraphics[width=\textwidth]{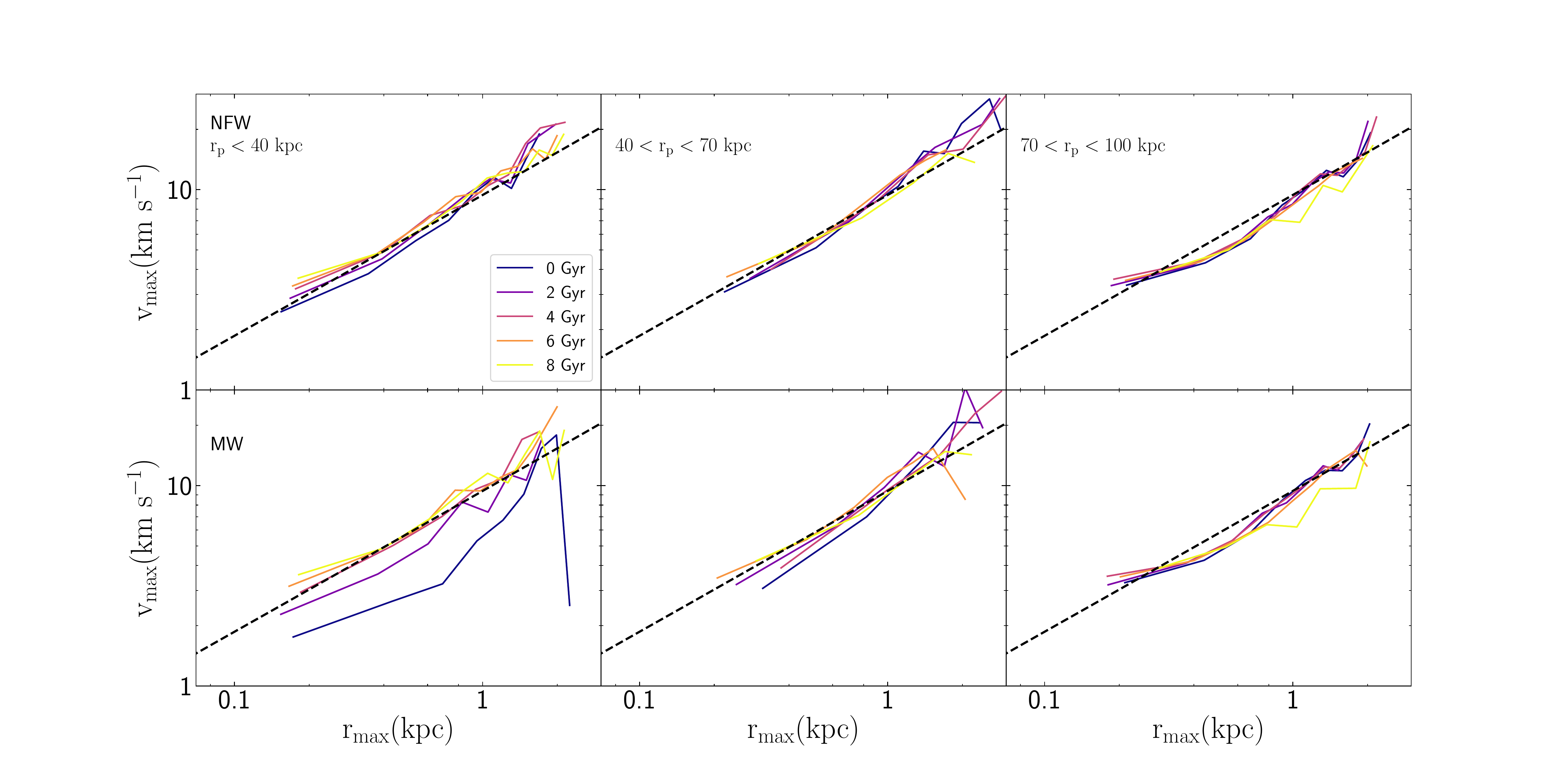}
    \caption{Radius of maximal circular velocity versus maximum circular velocity as a function of time (from 8 Gyr ago to present day (0 Gyr)) for resimulated VL2 subhalos with different pericentric radii(Left: $r_p$ $<$ 40 kpc, Middle: 40 $<$ $r_p$ $<$ 70 kpc, Right: 70 $<$ $r_p$ $<$ 100 kpc) . Top panels illustrate subhalos evolving in the NFW model and bottom panels illustrate subhalos evolving in the MW model. Uncertainties in the mean $\vmax$ are on the order of $1\,\mathrm{km\,s}^{-1}$, with the exception of the $t=0$ Gyr case where uncertainties reach $\sim 4\,\mathrm{km\,s}^{-1}$ at large $\rmax$ . The standard redshift zero CDM relationship from \citet{Dutton14a} is illustrated as a dashed line.}
   \label{fig:via_lactea_rvmax_rp}
\end{figure*}

If we again instead consider the relationship between $\rmax$ and $\vmax$ of subhalos in each galaxy model at different times and for subhalos with different pericentres, a trend is revealed. Figure \ref{fig:via_lactea_rvmax_rp} demonstrates that while the structural properties of most subhalos in both the NFW and MW potentials are in agreement with predictions from CDM, subhalos with pericentres that bring them within 40 kpc have values of $\vmax$ that are much lower than CDM would predict given their $\rmax$. Hence interactions with the disk and bulge are causing subhalos to lose mass at a near constant $\rmax$.

\subsection{Subhalo orbit distribution}\label{s_orbits}

\begin{figure}
    \includegraphics[width=0.48\textwidth]{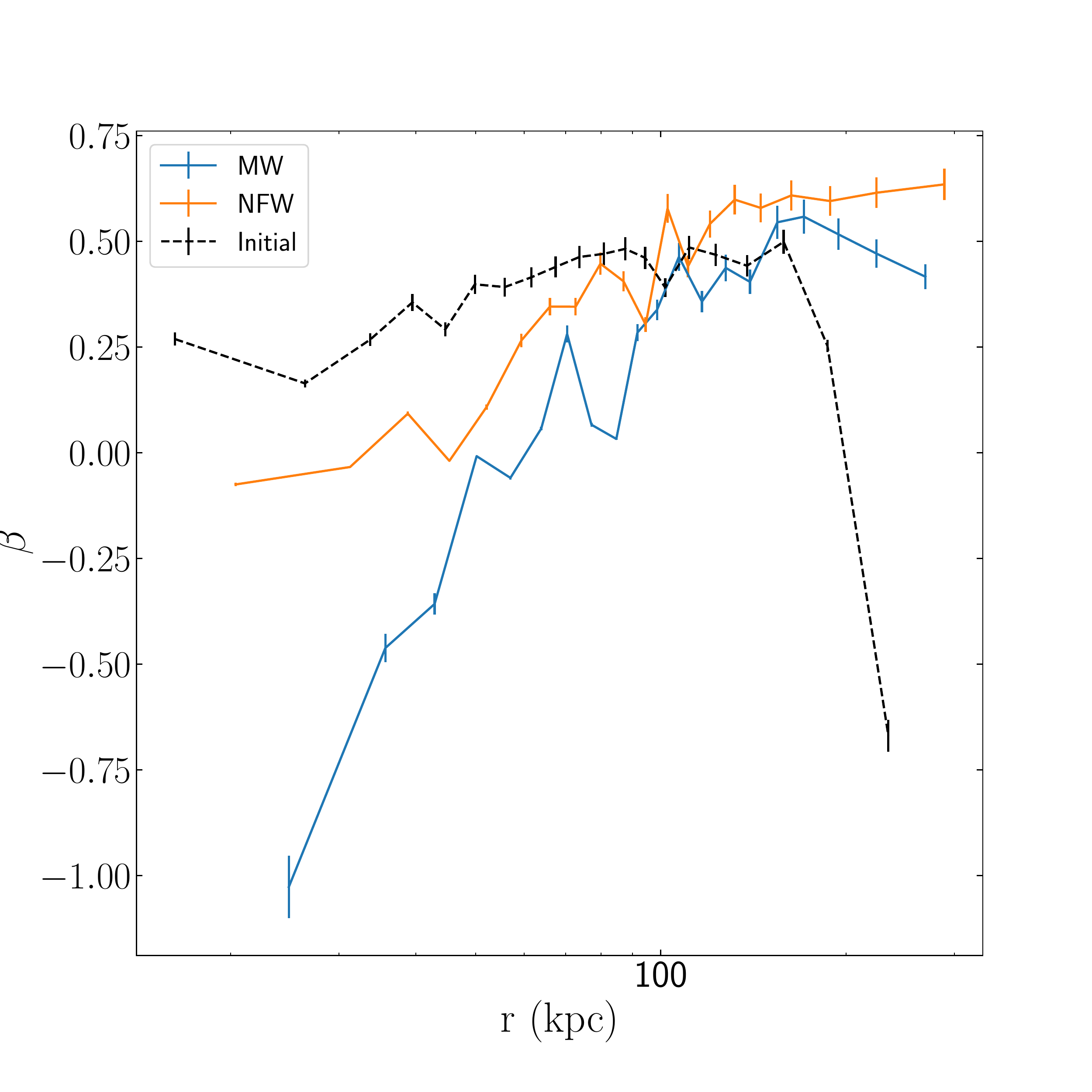}
    \caption{Orbital anisotropy profile of resimulated VL2 subhalos that survive until a redshift zero in the MW and NFW galaxy potentials. The profile of all subhalos at infall are marked is a dotted line. The errorbars represent Poisson error. Within 40\kpc, the subhalo distribution is much more tangentially anisotropic in the MW potential due to the preferential disruption of subhalos on radial orbits by the bulge and disk.}
   \label{fig:beta_prof}
\end{figure}

Finally, we examine the orbital anisotropy of the surviving subhalo population in each galaxy model. More specifically we investigate how the orbital anisotropy parameter $\beta$ changes as a function of instantaneous galactocentric radius $r$. Figure \ref{fig:beta_prof} illustrates that while the distribution of subhalo orbits at infall is relatively isotropic, the dissolution of inner region subhalos in both galaxy models leads to a tangential bias. Essentially inner region subhalos with eccentric orbits that bring them deep into the potential well of the galaxy quickly reach dissolution. The effect is less dramatic in the NFW model, with $\beta$ starting at $-1$ in the innermost regions of the galaxy and slowly climbing to being isotropic or slightly radially anisotropic in the outer regions of the galaxy. The difference between the infall anisotropy profile and the redshift zero profile in the NFW model is due to the combined effects of modelling the subhalos with higher resolution and having a static external tidal field. In the MW model, however, a wider range of inner region subhalos reach dissolution due to interactions with the bulge and disk. Hence $\beta$ starts at a much lower values ($-3$) and is fairly low out to 40 kpc. Beyond 40 kpc, it appears that the effect of the bulge and disk on subhalo evolution is diminished and the anisotropy profile more closely resembles that of the NFW potential, albeit still being somewhat more tangential. That the orbital distribution of surviving subhalos is highly tangential is because of the preferential disruption of subhalos on radial orbits by the bulge and disk. It is worth noting that the satellite galaxies of the Milky Way and simulated subhalos in AURIGA have anisotropy profiles that are similar to Figure \ref{fig:beta_prof} \citep{Riley19}.

\section{Discussion}

\subsection{Comparison to cosmological simulations}

This work marks the largest suite of high-resolution simulations of subhalo evolution in tidal fields containing dark matter-only and containing dark matter with an axisymmetric baryonic component to date. These simulations can be directly compared to lower resolution hydrodynamic cosmological simulations of galaxy formation to determine how poorer resolution affects estimates of how including baryonic matter influence the evolution of subhalos. Several studies discussed in Section \ref{s_intro} provide estimates for the ratio of the number of subhalos that survive to a redshift of zero in a Milky Way-like tidal field to the number that survive in a dark-matter only tidal field ($N_\mathrm{MW}/N_\mathrm{NFW}$). In order to evaluate how the results of this study compare to these works, it is necessary to first compare the mass resolution, softening length, and mass range used in each study.

As shown in Table \ref{simparams}, hydrodynamical cosmological simulations typically set the mass of individual dark matter particles to be between $10^{4}$ and $10^{6}$ $M_{\odot}$ with softening lengths on the order of 10 to 100 pc. It is important to note that \citet{wetzel16} and \citet{kelley19} provide their sub-halo range in terms of $\vmax$, for which we approximate a mass given the $\vmax$ and masses of subhalos in VL2. Our simulations, also listed in Table \ref{simparams}, have a mass resolution 1000 $\times$ better than these large-scale simulations and softening lengths that are 10 to 100 $\times$ smaller. Our choice of dark matter particle mass and softening length follow the criteria of \citet{vandenbosch18b} for evolving a subhalo to $1\%$ its an initial mass.

\begin{table*}
\centering
\begin{tabular}{@{}llcc@{}}
\hline\hline
Reference      & $M_\mathrm{DM}$ & Softening Length & Mass Range \\ 
{} & {$M_{\odot}$} & pc & {$M_{\odot}$} \\ \hline\hline
\citet{DOnghia10}       & $5.5\times10^{5}\, h^{-1}$    & $200\, h^{-1}$  & $> 10^{7}$  \\ \hline
\citet{wetzel16} (FIRE)         & $3.5\times10^{4}$     & 20  & $10^{6}$ to $10^{10}$  \\ \hline
\citet{Sawala17} (APOSTLE)  & $5.0\times10^{4}$,$5.9\times10^{5}$,$7.5\times10^{6}$     & 134, 307, 711 & $10^{6.5}$ to $10^{8.5}$  \\ \hline
\citet{kelley19} (ELVIS) & $3\times10^{4}$     & 37  & $10^{5.5}$ to $10^{10}$  \\ \hline
\citet{Richings20} (AURIGA)      & $10^{4}$     & 180  & $10^{6.5}$ to $10^{8.5}$  \\ \hline
This Work (Webb \& Bovy 2020)        & O(10)     & O(1)  & $10^{6}$ to $10^{7.5}$  \\

\hline\hline
\end{tabular}
\caption{The dark matter particle mass $M_\mathrm{DM}$ and softening length used in different cosmological simulations of galaxy formation that include dark and baryonic matter as well as the mass range over which the author's provide the equivalent of our $N_\mathrm{MW}/N_\mathrm{NFW}$ mass-function suppression factor. For reference purposes, the parameters used in this work are also provided where subhalos are simulated individually in an external tidal field containing dark and baryonic matter.}
\label{simparams}
\end{table*}

\begin{figure}
    \includegraphics[width=0.48\textwidth]{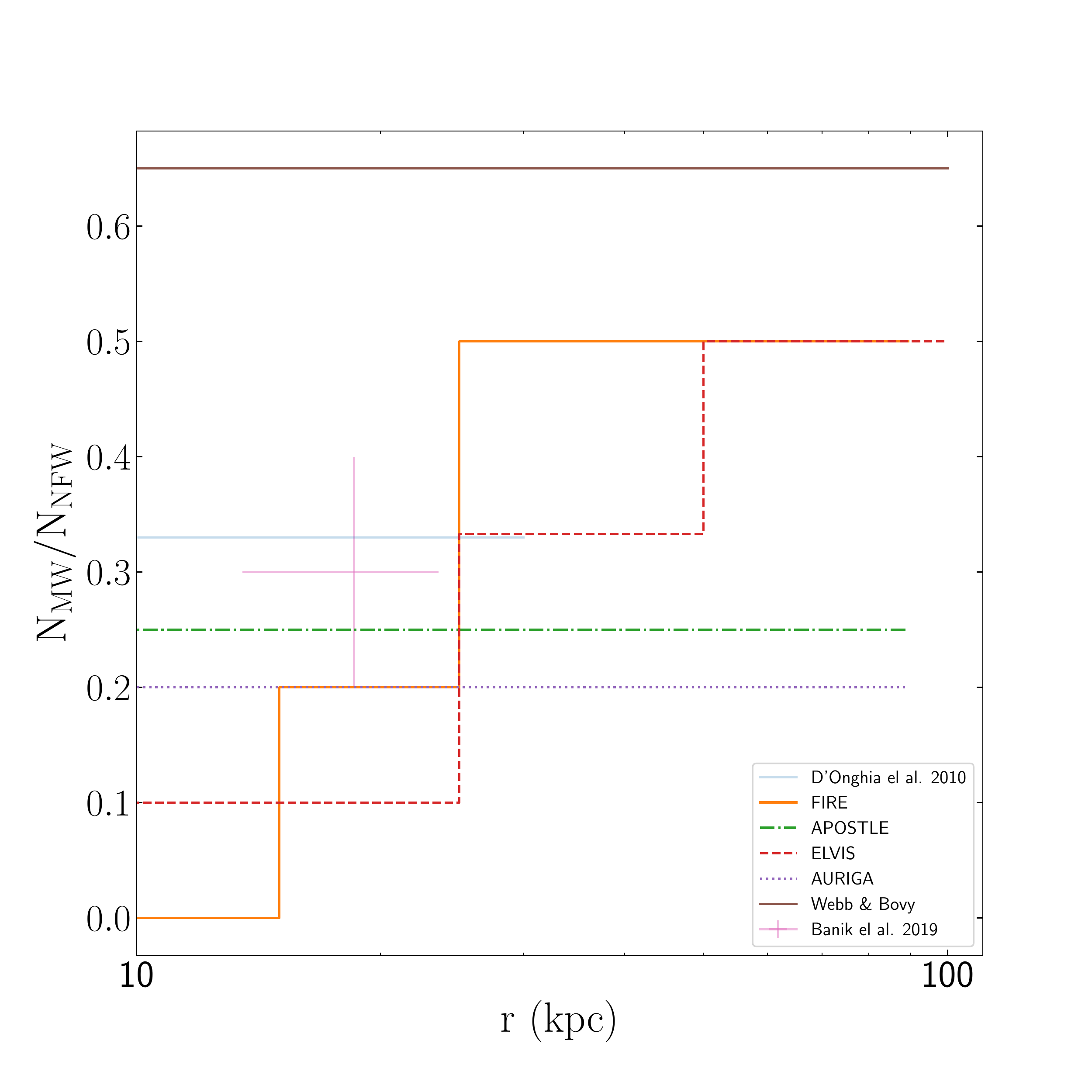}
    \caption{Ratio of number of subhalos in the MW potential to number of subhalos in dark-matter-only NFW potential as a function of galactocentric distance at redshfit of zero for subhalos with masses between $10^6$ and $10^{7.5}\,M_{\odot}$ in this work (Webb \& Bovy) compared to results from several previous studies. This study predicts a higher $N_\mathrm{MW}/N_\mathrm{NFW}$ than previous works at all galactocentric radii.}
   \label{fig:res_compare}
\end{figure}

Each of the studies listed in Table \ref{simparams} provide estimates for $N_\mathrm{MW}/N_\mathrm{NFW}$ over different radial ranges, which are illustrated in Figure \ref{fig:res_compare}. It is important to note that each study considers a slightly different range of subhalo masses when calculating $N_\mathrm{MW}/N_\mathrm{NFW}$, so a direct comparison is not entirely straightforward. For comparison purposes, the constraint placed on $N_\mathrm{MW}/N_\mathrm{NFW}$ by \citet{banik19b} within the orbits of the GD-1 and Pal 5 streams is illustrated in Figure \ref{fig:res_compare}. In total we find that $\Lambda$CDM overestimates the number of subhalos in the galaxy by a factor of $\sim 65\%$ at all radii, while cosmological simulations estimate a much higher substructure disruption rate due to baryonic physics. The discrepancy is larger in the inner regions of the galaxy model due to the baryonic disk being significantly more denser than in the outer parts of the galaxy model.

The study with the smallest discrepancy with our work in the inner regions is \citet{DOnghia10}, which at first is surprising given their large dark matter particle mass and softening length. However taking into consideration that this study focuses on higher mass subhalos only, the lack of subhalo disruption is understandable as high mass subhalos will have longer dissolution times than low-mass subhalos. The simulations with the largest discrepancy from our work are FIRE and ELVIS. \citet{wetzel16} finds there to be no subhalos within 15 kpc of the Galactic centre in the FIRE simulations while \citet{kelley19} finds a factor of 5 fewer subhalos in the inner regions of the ELVIS simulations. Other works estimate closer to a factor of between 2 to 3 fewer subhalos than this work. Some discrepancy is expected, since our galaxy model is static with time while in cosmological simulations the galaxy grows with time. However not accounting for galaxy growth should result in our study overestimating subhalo disruption, which is the opposite of what we find when comparing our work to cosmological simulations.

In the outer regions of the galaxy models, where baryonic physics is less important, we are in closer agreement ($\sim < 5\%$) with cosmological simulations like FIRE and ELVIS. However we still predict a factor of 2-3 more subhalos than APOSTLE or AURIGA. The overarching trend in the outer regions, where baryonic physics is less of a factor, is that cosmological simulations with lower mass resolution and larger softening lengths estimate higher subhalo disruption rates than our high-resolution simulations, in agreement with \citet{vandenbosch18a} and \citet{vandenbosch18b}.

In order to confirm the above statement about lower mass resolution and larger softening lengths leading to higher subhalo disruption rates we re-simulate a subset of our high-resolution subhalos in the NFW galaxy model with the mass resolution and softening lengths of the FIRE simulations (which we label as being low-resolution). More specifically, we randomly selected 300 subhalos with initial masses greater than $10^6 M_{\odot}$ for re-simulation and compare the results to our initial high-resolution simulations. Below $10^6 M_{\odot}$, low-resolution subhalos are initially made up of less than 100 dark matter particles and dissolve almost immediately. 

\begin{figure}
    \includegraphics[width=0.48\textwidth]{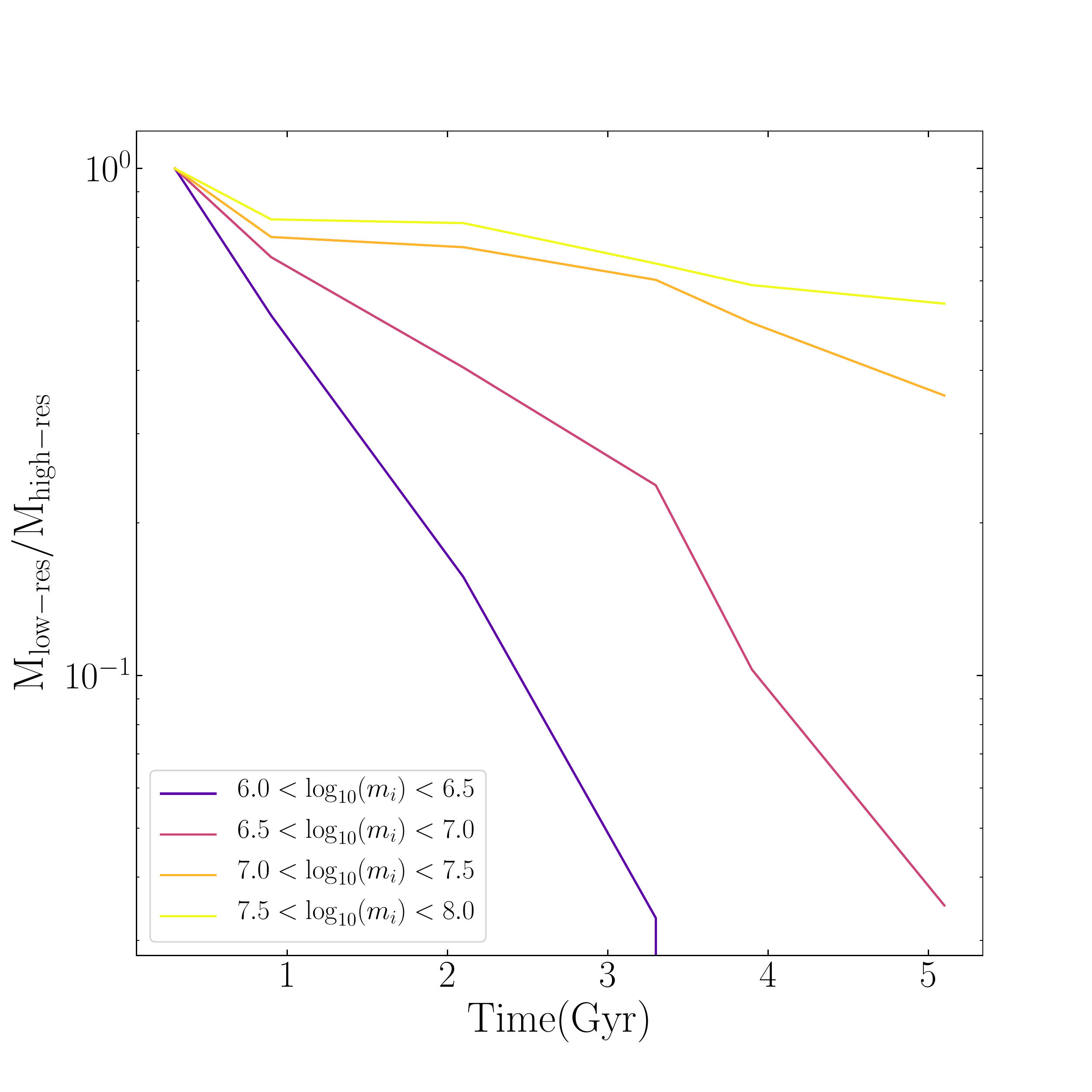}
    \caption{Ratio of total subhalo mass when simulating using the dark matter particle masses and softening lengths of the FIRE simulations (low-res) to the total mass of the same subhalo simulating in this work (high-res). High dark matter particle masses and larger softening lengths in the low-resolution simulation lead to higher mass-loss rates and  artificially shorter subhalo dissolution times.}
   \label{fig:subhalo_res_compare}
\end{figure}

The average mass evolution $\rm M_{low-res}$ of subhalos with different intial masses simulated using the dark matter particle mass and softening length of the FIRE simulations, normalized by the mass evolution of the same subhalo in our high-resolution simulations $\rm M_{high-res}$, are illustrated in Figure \ref{fig:subhalo_res_compare}. The average mass evolution is found by binning the simulations based on initial sub-halo mass and then determining the average ratio of $\rm M_{low-res}/M_{high-res}$ as a function of time. As expected, subhalos experiences a higher mass loss rate when evolved with parameters from FIRE. For subhalos with initial masses greater than or equal to $10^{7} M_{\odot}$, the low-resolution simulations can differ from the high-resolution simulations by up to $60\%$ over the course of their lifetimes. For subhalos with initial masses less than $10^{7}\,M_{\odot}$, low-resolution subhalos are able to dissolve completely while high resolution subhalos remain in-tact. Hence the differences in predicted values of $N_\mathrm{MW}/N_\mathrm{NFW}$ between our work and cosmological simulations can be attributed to the significant differences in resolution. 

\subsection{Implications for subhalo searches}

We predict nearly a factor of two \textit{fewer} subhalos in a Milky Way-like galaxy model than the standard $\Lambda$CDM cosmological model. The immediate implication of this conclusion is that studies focused on detecting subhalos will have lower detection rates than naively expected from dark-matter-only CDM predictions. For example, interaction rates between stellar streams and subhalos and substructure fractions in gravitational lenses will be significantly smaller than those estimated using $\Lambda$CDM.

We predict a factor of two fewer subhalos in a Milky-Way like galaxy with little dependence on subhalo mass or present-day galactocentric radius (despite a strong dependence on \emph{pericentre} radius, washed out because of the mix of different pericentres at any given radius). Thus, subhalo abundance measurements like those of \citet{banik19b} should find a suppressed mass function at all radii. It is interesting to note that we predict a factor of two \textit{more} subhalos in a Milky Way-like galaxy model than the observational constraint of \citet{banik19b}, who uses the GD-1 and Pal5 streams to estimate $0.2 < N_\mathrm{MW}/N_\mathrm{NFW} <0.4$. In fact the works of \citet{DOnghia10} and \citet{Sawala17} are in better agreement with \citet{banik19b}. However, this apparent agreement is likely coincidental as these studies focus on slightly higher mass subhalos than \citet{banik19b} and are affected by accelerated subhalo disruption rates due to poor mass resolution and force softening (see above).

Several factors known to disrupt globular clusters, like tidal shocks due to baryonic substructure \citep[e.g.][]{Gieles06, Gieles07} and galaxy growth via hierarchical mergers \citep[e.g.][]{Kruijssen11}, are not considered here. Subhalo-subhalo interactions, while shown to have a minor effect on cluster evolution \citep{webb19}, may also prove crucial to subhalo evolution, because subhalos are significant more extended than globular clusters (and therefore have a larger interaction cross-section). While \citet{garrisonkimmel17} concluded that subhalo evolution was similar in galaxy models where the baryonic component was either analytic or made up of live baryonic particles, if subhalo dissolution is dominated by their mass resolution and softening length, it is possible the importance of a more detailed baryonic potential was suppressed. Including these additional factors may minimize the factor of two difference between our work and \citet{banik19b}.

Dark matter substructure detection via gravitational lensing will also be affected by tidal disruption, although less than estimated here, because strong gravitational-lensing studies typically focus on elliptical galaxies, with a weaker baryonic tidal field \citep[e.g.][]{gilman20}. Additionally, gravitational lensing is sensitive to substructure along the line of sight that is unassociated with galaxies (and thus, likely unsuppressed) and is affected by subhalos within the entire line-of-sight cylinder through the lens, whose volume is dominated by large galactocentric radii. Thus, for gravitational lenses we expect $N_\mathrm{MW}/N_\mathrm{NFW} > 0.65$. However, resolution effects may affect the estimated tidal-disruption rate in strong gravitational lenses and similar studies such as the one performed here may be warranted for making detailed predictions for the expected CDM substructure fraction in gravitational lenses.

\section{Conclusion}\label{s_conclusion}

Through a large suite of high-resolution dark matter subhalo simulations, we explore how subhalos evolve in galaxy models with and without a stellar bulge and an axisymmetric stellar disk. For each individual subhalo, the dark matter subhalo mass and softening length are set so the dissolution of the subhalo can be followed down to $1\%$ of its initial mass. These simulations suggest that including a baryonic componenent in the external tidal field leads to more subhalo dissolution overall, with surviving subhalos typically be less massive than if they evolved in a dark matter-only tidal field.

Based on our suite of simulations, we predict that not including the effects of baryonic matter results in $\Lambda CDM$ overestimating the number of subhalos within the Milky Way's virial radius by a factor of $\approx 1.6$, independent of galactocentric distance (that is, the expected subhalo mass function is $\approx 65\,\%$ of that predicted from dark-matter-only simulations). Taking into consideration each subhalo's orbit, subhalos with small pericentres that bring them close to the bulge or axisymmetric disk are more strongly affected than subhalos that orbit primarily in the outer regions of the galaxy. More specifically, the number of surviving subhalos with pericentres less than 40 kpc in the MW model is $40\%$ of what is found in the NFW model. For subhalos with larger pericentres, the ratio is only $70$ to $80\%$, because there is minimal interaction between outer subhalos and baryonic matter.

Given how strongly including axisymmetric stellar components in the external potential affects subhalo evolution, we also expect the surviving subhalo population to have different structural and orbital properties. Looking closer at the structural properties of the surviving subhalos, we find that subhalos that evolve in a realistic, Milky Way-like potential have a lower density because the shape of their density profile remains constant while they lose mass. Surviving subhalos also have a strong orbital anisotropy gradient. More specifically, inner subhalos only survive to reach a redshift of zero if their orbits are primarily circular such that they are not brought deep into the central regions of the galaxy. Hence we find inner subhalos to be tangentially anisotropic while outer subhalos, which are only weakly affected by including an axisymmetric stellar component in the external potential, have an isotropic distribution of orbits.

When comparing to previous large-scale cosmological simulations like FIRE and ELVIS, we predict factors of between 2 and 5 more subhalos in the inner regions of Milky Way-like galaxies than these simulations claim. In the outer regions of the galaxy models, where the baryonic components of the potential are weak, we are in much closer agreement with these large-scale simulations. We confirm that discrepancies between our high-resolution simulations of subhalo evolution and previous estimates from large-scale cosmological simulations can be attributed to the higher dark matter particle mass and longer softening lengths used in these simulations leading to accelerated dissolution times (see \citealt{vandenbosch18a} and \citealt{vandenbosch18b} for a thorough investigation of this issue). 

When comparing to observations, we also estimate there to be more subhalos in Milky Way-like galaxies than \citet{banik19b} predicts using the GD-1 and Pal 5 streams, by a factor of between $\sim$ 1.4 and 1.8.  Given that GD-1 and Pal 5 have Galacotocentric distances of $\sim 14$ kpc and $\sim 18.5$ kpc respectively, it is worth noting that \citet{banik19b}'s estimate is consistent with the behaviour of subhalos in our simulations with pericentres less than 25 kpc. 

Ultimately, we conclude that including the effects of baryonic matter is necessary when estimating how subhalos evolve in Milky Way-like galaxies. Furthermore, dark matter particle masses that are too high or softening lengths that are too large can lead to the dissolution of individual subhalos being artificially accelerated. Only when both these factors are considered can the properties of dark matter subhalo populations in simulations be compared to observations. The next step will be to consider whether or not a non-static external tidal field could potentially minimize the gap between theoretical and observational constrains placed on dark matter substructure. Such a step is an important one to make given recent advancements in stream finding and gravitational lensing studies, both of which are commonly used to indirectly measure the presence and properties of dark matter substructure.

\section*{Acknowledgements}

JB acknowledges financial support from NSERC (funding reference number RGPIN-2015-05235) and an Ontario Early Researcher Award (ER16-12-061).

\section*{Data Availability}

Data from the Via Lactea II (VL2) simulation \citep{diemand07, diemand08}, which was used to generate the initial conditions of our subhalo simulations, is publicly available at 
https://www.ucolick.org/~diemand/vl/. The large suite of simulations discussed in this article, as well as Notebooks for their setup and analysis,  will be shared on reasonable request to the corresponding author.




\bibliographystyle{mnras}
\bibliography{ref2} 


\bsp	
\label{lastpage}
\end{document}